\documentclass[aps,letterpaper]{revtex4}
\usepackage{amsmath,amssymb,amsfonts} 
\usepackage{color,graphicx}
\usepackage{multirow}

\def\bq{\begin{equation}}
\def\eq{\end{equation}}
\def\ba{\begin{eqnarray}}
\def\ea{\end{eqnarray}}

\begin{document}

\title{Plasmonic Cloaking of Cylinders: \\
       Finite Length, Oblique Illumination
       and Cross-Polarization Coupling}

\author{Andrea Al\`u\footnote{corresponding author}}
  \email{alu@mail.utexas.edu}
  \affiliation{Dept.\ of Electrical and Computer Engineering,
               The University of Texas at Austin,
               Austin, TX 78712, USA}

\author{David Rainwater}
  \email{rain@arlut.utexas.edu}
  \affiliation{Applied Research Laboratories,
               The University of Texas at Austin,
               Austin, TX 78758-4423, USA}

\author{Aaron Kerkhoff}
  \email{kerkhoff@arlut.utexas.edu}
  \affiliation{Applied Research Laboratories,
               The University of Texas at Austin,
               Austin, TX 78758-4423, USA}

\begin{abstract}
  Metamaterial cloaking has been proposed and studied in recent years
  following several interesting approaches.  One of them, the
  scattering-cancellation technique, or plasmonic cloaking, exploits
  the plasmonic effects of suitably designed thin homogeneous
  metamaterial covers to drastically suppress the scattering of
  moderately sized objects within specific frequency ranges of
  interest.  Besides its inherent simplicity, this technique also
  holds the promise of isotropic response and weak polarization
  dependence.  Its theory has been applied extensively to symmetrical
  geometries and canonical 3D shapes, but its application to elongated
  objects has not been explored with the same level of detail.  We
  derive here closed-form theoretical formulas for infinite cylinders
  under arbitrary wave incidence, and validate their performance with
  full-wave numerical simulations, also considering the effects of
  finite lengths and truncation effects in cylindrical objects.  In
  particular, we find that a single isotropic (idealized) cloaking
  layer may successfully suppress the dominant scattering coefficients
  of moderately thin elongated objects, even for finite lengths
  comparable with the incident wavelength, providing a weak dependence
  on the incidence angle.  These results may pave the way for
  application of plasmonic cloaking in a variety of practical
  scenarios of interest.
\end{abstract}

\date{\today}

\maketitle


\vspace*{-5mm}

\section{Introduction}

The application of metamaterials to cloaking and invisibility has been
explored in several exciting papers in recent
years~\cite{Pendry2006,Schurig2006,Cai2007,Leonhardt2006,Milton2005,Milton2006,Greenleaf2007,Miller2006,Yaghjian2008,Alitalo2008,Tretyakov2009,Castaldi2009,Lai2009a,Lai2009b,Alu2005a}.
In these contributions, several alternative realizations and
techniques have been discussed, and the exotic properties of several
classes of metamaterials have been shown to be possibly tailored in order to provide much reduced scattering from finite-sized objects in different
configurations and schemes, for a wide range of frequencies of
interest.  Recent reviews of the various possibilities are available
for the interested reader (see e.g.\
Refs.~\cite{Alitalo2009,Alu2008a}).

One such possibility takes advantage of the anomalous scattering
response of thin plasmonic layers.  As shown in Ref.~\cite{Alu2005a},
artificial plasmonic materials with low or negative effective
permittivity may provide scattering cancellation via their local
negative polarizability.  This technique, called plasmonic cloaking,
is consistent with earlier works that have speculated how a composite
particle combining positive and negative permittivity may provide
identically zero scattering in the static limit~\cite{Kerker1975}.
Recent study of the dynamic case has shown that plasmonic cloaks may
suppress not only the dominant dipolar scattering from moderately
sized objects, but also higher-order multipolar orders for larger
scatterers~\cite{Alu2008b}.  In this vein, it is worth stressing that
by ``cloaking'' we mean strong, or maximized, scattering reduction
over a finite frequency band -- not necessarily complete invisibility,
since residual scattering orders may always make the scatterer
detectable to a certain extent.  Still, significant reduction of
visibility is achievable with the proper plasmonic cloak design, as
shown in several recent
papers~\cite{Alu2007a,Alu2008c,Alu2008d,Alu2007b,Alu2008e,Alu2008f,Alu2009a,Alu2009b,Tricarico2010}
and as discussed in the following.

Plasmonic cloaking has been shown to offer several intriguing
properties in a variety of setups.  Examples include intrinsic
robustness to frequency and design
variations~\cite{Alu2007a,Alu2008c,Alu2008d}, straightforward
extension to arbitrary collections of objects~\cite{Alu2007b} and
multi-frequency operation~\cite{Alu2008e,Alu2008f}.  Moreover, the
admission of fields inside the cloaked region, peculiar to this
technique, may be used to suppress the inherent scattering from
receiving antennas and sensing devices, which may open several
interesting venues in non-invasive probing and sensing
applications~\cite{Alu2009a}.  Extensions to ultrathin surface cloaks
have also been put forward in this same context~\cite{Alu2009b}.  All
these studies were conducted, for simplicity, for a canonical
spherical object, illuminated by a generic polarization of the
impinging field.  Recent studies~\cite{Tricarico2010} have shown that
analogous concepts may be applied to more complex 3D geometries, based
on the overall robustness of the integral cancellation effect.

However, in several practical applications, in particular for the
radar community, elongated objects may become of specific interest for
these applications.  Ref.~\cite{Alu2005a} provides quasi-static
formulas for infinite circular dielectric cylinders under normal
incidence, which was later extended to 2D infinite conducting
cylinders, also illuminated at normal incidence in
Ref.~\cite{Irci2007}.  These results were also preliminarily extended
to oblique incidence in Ref.~\cite{Irci2007th}.  Moreover, in
Ref.~\cite{Wang2009} the results of Ref.~\cite{Alu2008e} for
multi-frequency operation of spherical cloaks were extended to
dielectric infinite cylinders.  Also, recent theoretical and
experimental efforts~\cite{Silveirinha2007,Edwards2009} reported the
practical realization of plasmonic cloaking in 2D cylindrical
geometries, proposing metamaterial designs for specific polarization
of interest.

The literature on plasmonic cloaking applied to cylinders, however,
has often dealt with idealized 2D geometries: infinite cylinders, and
incident waves normal to the cylinder axis, with specific polarization
properties.  This assumption, common to several other cloaking
techniques applied to cylinders, makes the resulting calculations
quite limited from a practical standpoint.  It may be argued, in
particular, that once the angle of incidence is modified, and the end
effects of finite-length cylinders are considered, such cloaking
effects may be severely limited, if not completely lost.  In
~\cite{Bilotti2008,Bilotti2010}, the effects of truncation were
preliminarily considered for normal incidence.

Here we analyze all these issues in great detail, first deriving a
general cloaking theory for arbitrary illumination of an infinite
cylinder.  We show that it is indeed possible to find a suitable,
robust plasmonic cloaking layer in this scenario, which under suitable
conditions may operate over a broad range of incidence angles.  We
corroborate these findings with an extensive numerical analysis of
finite length and truncation effects in practical cylindrical
geometries, studying the overall scattering reduction for different
angles of incidence and several design parameters.  The present
analysis applies to idealized metamaterial cloaks with isotropic
properties.  We leave to future work the practical limitations
introduced by the specific realization of engineered metamaterial
cloaks.


\section{Analytical Results for Infinite Cylinders}
\label{sec:anal}


\subsection{General Derivation}

\begin{figure}[!t]
\centering
\includegraphics[width=4.7cm]{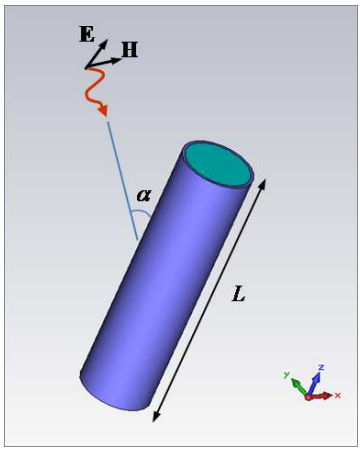}
\vspace*{-3mm}
\caption{Schematic of the cloaked cylindrical object problem and its
  geometry.  A dielectric core of length $L$ and diameter $2a$ is
  covered uniformly, except the ends, by a metamaterial shell of
  radial thickness ($a_c-a$).  The object is illuminated by an arbitrarily polarized plane
  wave, incident at an angle $\alpha$ with respect to the cylinder axis.
}
\label{fig:cyl}
\vspace*{-2mm}
\end{figure}

Consider the geometry of Fig.~\ref{fig:cyl}, depicting a circular
cylinder of finite length $L$, radius $a$, permittivity $\epsilon$ and
permeability $\mu$, covered by a thin conformal cylindrical cloak
shell of thickness $(a_c-a)$, permittivity $\epsilon_c$ and
permeability $\mu_c$.  In this section, we examine the limiting case
of an infinite circular cylinder ($L\to\infty$) illuminated by an
arbitrary plane wave at oblique angle.  The general scattering problem
may be solved by expanding the impinging and scattered fields in
cylindrical harmonics (see e.g.
Refs.~\cite{Balanis1989,Bohren1983,Yousif1994}), which we apply to
derive the scattering response as a function of incidence angle
$\alpha$.

Without loss of generality, the incident wave may be decomposed into
its transverse-magnetic (TM$_z$) and transverse-electric (TE$_z$)
components, with respect to the cylinder ($z$) axis. By matching
boundary conditions at the two radial interfaces, the problem may be
solved exactly for each cylindrical harmonic
~\cite{Balanis1989,Bohren1983,Yousif1994}.  Analytical results for
normal incidence, specifically applied to the cloaking problem at
hand, were derived in Ref.~\cite{Silveirinha2007}.  In that special
case ($\alpha=90^\circ$), only two tangential field components are
non-zero at each interface: $E_z,H_\phi$ for TM polarization, and
$E_\phi,H_z$ for TE.  This ensures that the scattering problem is
easily solved as a set of four equations, and the TE and TM Mie
scattering coefficients are easily expressed as a function of
$4\times4$ determinants, $U_n^{\rm TM,TE}$ and $V_n^{\rm
  TM,TE}$~\cite{Silveirinha2007}:

\bq\label{eq:Mie}
c_n \, = \, \frac{-U_n}{U_n+iV_n}
\eq
for either TM or TE polarization. We have assumed here and in the
following an $e^{-i\omega t}$ time dependence.  The overall scattering
efficiency, defined as the ratio of the total scattering cross section
normalized to the cylinder's physical cross section, is given
by~\cite{Bohren1983}:
\bq\label{eq:sum}
C_s \, = \, \frac{2}{k_0 a_c} \sum_{n=-\infty}^{\infty}
\biggl( |c_n^{\rm TM}|^2 + |c_n^{\rm TE}|^2 \biggr) \, .
\eq
As discussed in ~\cite{Silveirinha2007}, by tuning the cloak's
electrical and geometrical properties, the cylinder's visibility can
be minimized by canceling the dominant scattering orders, which is
possible under the condition $U_n=0$.  In the quasi-static limit,
i.e., for very thin cylinders, approximate closed-form expressions for
these cloaking conditions for TE and TM incidence were reported in
Refs.~\cite{Alu2005b,Irci2007}.

Scattering for oblique incidence is a more challenging problem, since
polarization coupling is usually involved~\cite{Yousif1994}, as
discussed in the Appendix~\ref{app}.  This implies that for oblique
incidence purely TE or TM cylindrical waves may excite also TM and TE
scattered waves, respectively.  This feature will play an important role
when analyzing the response of the cloak for different incidence
angles.  In the special circumstance for which this coupling may be
absent, or minimized, e.g., as discussed in the Appendix, for azimuthally symmetric modes
($n=0$), for conducting or high-contrast cylinders, or for small
cross-sections, the scattering coefficients may be still expressed as
in Eq. (1), where $U_n$ and $V_n$ have generalized expressions:

\bq \label{eq:Un}
U_n^{\rm TM} \, = \,
\begin{array}{|cccc|}
J_n(k^T a)   &
J_n(k_c^T a) &
Y_n(k_c^T a) &
0            \\
\frac{k  }{\eta   k^T  } J_n^\prime(k^T a)   &
\frac{k_c}{\eta_c k_c^T} J_n^\prime(k_c^T a) &
\frac{k_c}{\eta_c k_c^T} Y_n^\prime(k_c^T a) &
0                                            \\
0 &
J_n(k_c^T a_c) &
Y_n(k_c^T a_c) &
J_n(k_0^T a_c) \\
0 &
\frac{k_c}{\eta_c k_c^T} J_n^\prime(k^T a_c)   &
\frac{k_c}{\eta_c k_c^T} J_n^\prime(k_c^T a_c) &
\frac{k_0}{\eta_0 k_0^T} J_n^\prime(k_c^T a_c)
\end{array}
\eq
and
\bq \label{eq:Vn}
V_n^{\rm TM} \, = \,
\begin{array}{|cccc|}
J_n(k^T a)   &
J_n(k_c^T a) &
Y_n(k_c^T a) &
0            \\
\frac{k  }{\eta   k^T  } J_n^\prime(k^T a)   &
\frac{k_c}{\eta_c k_c^T} J_n^\prime(k_c^T a) &
\frac{k_c}{\eta_c k_c^T} Y_n^\prime(k_c^T a) &
0                                            \\
0 &
J_n(k_c^T a_c) &
Y_n(k_c^T a_c) &
Y_n(k_0^T a_c) \\
0 &
\frac{k_c}{\eta_c k_c^T} J_n^\prime(k^T a_c)   &
\frac{k_c}{\eta_c k_c^T} J_n^\prime(k_c^T a_c) &
\frac{k_0}{\eta_0 k_0^T} Y_n^\prime(k_c^T a_c)
\end{array}
\; ,
\eq
\\
where $J_n$ and $Y_n$ are the cylindrical Bessel functions of the
first and second kind of order $n$, $J^\prime,Y^\prime$ are their
derivatives with respect to the argument, and
$\eta_i=\sqrt{\mu_i/\epsilon_i}$ are the characteristic impedances of
each region.  The TE expressions may be found by electromagnetic
duality: $\epsilon_i\leftrightarrow\mu_i$.

These expressions coincide with those in Ref.~\cite{Silveirinha2007}
for normal incidence ($\alpha=90^\circ$), as expected.  The dependence
on the angle of incidence $\alpha$ is encoded in the transverse
wavenumbers, $k_i^T$.  Specifically, $k_i^T=\sqrt{k_i^2-\beta^2}$,
where $\beta=k_0\cos\alpha$ is the wave number component along the
cylinder axis.  In particular, $ k_0^T =k_0\sin\alpha$, as expected.
In the general case, where polarization coupling will occur,
Eq.~\ref{eq:Mie} should be modified to include the TM--TE coupling
coefficients, consistent with the derivation in the Appendix and the
results reported in [36].


\subsection{Quasi-Static Analysis}

As in the spherical scattering scenario~\cite{Alu2005a} and
cylindrical scenario at normal incidence~\cite{Silveirinha2007}, it is
instructive to first consider the case of electrically small
cylinders, i.e.\ $k_i a_c\ll 1$.  In this limit, as we discuss in the
following, the analysis is made simpler by the fact that: (a) the
dependence of Eqs.~(\ref{eq:Un},\ref{eq:Vn}) on the incidence angle is
negligible; consequently, so is the cross-polarization effect on the
cloaking conditions; (b) the scattered wave is dominated by a limited
number of multipolar orders. The combination of these features makes
the plasmonic cloaking easier to achieve, more effective and robust,
compared to larger cylinders.  This is expected, since satisfying the
condition $U_n=0$ for only few dominant terms in Eq.~(\ref{eq:sum}) is
easier to achieve, and such conditions are less dependent on the angle
of incidence in this regime.  In particular, it is easy to show that
for regular dielectric cylinders ($\epsilon>\epsilon_0$, $\mu=\mu_0$),
the dominant terms in Eq.~(\ref{eq:sum}) are $c_0^{\rm TM}$ and
$c_1^{\rm TE}$, respectively for TM and TE excitations.  For arbitrary
polarization, $c_0^{\rm TM}$ scattering dominates.  This also applies
to a conducting cylinder, whose limiting case corresponds to
$\epsilon\to-j\infty$ and $\mu\to0$ in Eqs.~(\ref{eq:Un},\ref{eq:Vn}).
For a magnetic cylinder ($\mu>\mu_0$, $\epsilon=\epsilon_0$) dual
considerations apply, so $c_0^{\rm TE}$ dominates.

\subsubsection{Quasi-static cloaking conditions}

As discussed in the Appendix, for electrically small cylinders ($k_i
a_c\ll 1$), the cross-coupling terms vanish for any angular order $n$.
We may thus derive the quasi-static cloaking condition by simply
taking the first-order Taylor expansion of the coefficients of
Eq.~(\ref{eq:Mie}).  This leads to simple closed-form solutions for
the appropriate ratio of cloak to core radii to achieve cloaking in
the quasi-static limit for each scattering coefficient, consistent
with analogous formulas available in the spherical~\cite{Alu2005a} and
cylindrical normal-incidence~\cite{Silveirinha2007} scenarios:
\begin{align}\notag
c_0^{\rm TE} \, : \quad \frac{a_c}{a} \; = \; &
\sqrt{\frac{\mu_c-\mu}{\mu_c-\mu_0}}
\\ \notag
c_{n\ne0}^{\rm TE} \, : \quad \frac{a_c}{a} \; = \; &
\sqrt[2n]{\frac{(\epsilon_c-\epsilon)(\epsilon_c+\epsilon_0)}
               {(\epsilon_c-\epsilon_0)(\epsilon_c+\epsilon)}}
\\ \label{eq:FOT-gen}
c_0^{\rm TM} \, : \quad \frac{a_c}{a} \; = \; &
\sqrt{\frac{\epsilon_c-\epsilon}{\epsilon_c-\epsilon_0}}
\\ \notag
c_{n\ne0}^{\rm TM} \, : \quad \frac{a_c}{a} \; = \; &
\sqrt[2n]{\frac{(\mu_c-\mu)(\mu_c+\mu_0)}
               {(\mu_c-\mu_0)(\mu_c+\mu)}}
\end{align}
For perfectly electric conducting (PEC) cylinders, a case of
particular interest for cloaking applications at radio frequencies, we
find:
\begin{align}\notag
c_0^{\rm TE} \, : \quad \frac{a_c}{a} \; = \; &
\sqrt{\frac{\mu_c}{\mu_c-\mu_0}}
\\ \label{eq:FOT-PEC}
c_{n\ne0}^{\rm TE} \, : \quad \frac{a_c}{a} \; = \; &
\sqrt[2n]{\frac{\epsilon_c+\epsilon_0}
               {\epsilon_0-\epsilon_c}}
\\ \notag
c_{n\ne0}^{\rm TM} \, : \quad \frac{a_c}{a} \; = \; &
\sqrt[2n]{\frac{\mu_c+\mu_0}{\mu_c-\mu_0}}
\end{align}
We note that Ref.~\cite{Irci2007} misreported that there is no
possible quasi-static condition for $c_0^{\rm TE}$.  The correct
formula, shown above, results from taking the proper limits in our
previous analysis for $\epsilon\to-j\infty$ and $\mu\to0$ in $U_n^{\rm
  TE}$.  This is particularly relevant for oblique incidence, since
enforcing only $\epsilon\to-j\infty$ to model a PEC boundary, as
suggested in Ref.~\cite{Irci2007}, would not ensure zero normal
component of the magnetic field on the boundary, as required in the
PEC limit~\cite{Sihvola2009}.  As correctly reported
in~\cite{Irci2007}, there is no quasi-static condition for the
$c_0^{\rm TM}$ coefficient in the PEC scenario.

A few important points should be highlighted regarding
Eqs.~(\ref{eq:FOT-gen}-\ref{eq:FOT-PEC}).  First of all,
Eq.~(\ref{eq:FOT-gen}) for TE polarization formally coincides with the
derivation of Ref.~\cite{Alu2005a}.  The formulas here have been
obtained for completely generic oblique incidence, generalizing the
previously published result.  They show that in the long wavelength
limit, cloaking conditions are {\it unaffected by an arbitrary
  variation of the angle of incidence}.  In this regime, the TM--TE
polarization cross terms are of second-order (cf. Appendix~\ref{app}),
thus Eq.~(\ref{eq:FOT-gen}) ensures that an electrically small
magnetodielectric cylinder may always be cloaked using its dominant
scattering order in the Taylor expansion, regardless of the angle of
incidence.  As in the spherical scenario, these quasi-static formulas
split the role of permittivity and permeability between TM and TE
polarizations.  Special attention should be paid to the azimuthally
symmetric modes, for which the role of permittivity or permeability is
reversed compared to higher-order modes.  Moreover, again as in the
spherical case, the cloaking conditions depend only on the ratio
$a_c/a$, implying that a thin homogeneous shell may be employed to
suppress the relevant scattering orders in this quasi-static scenario.
It should be stressed that Eqs.~(\ref{eq:FOT-gen}-\ref{eq:FOT-PEC})
may not be met by arbitrary values of the constituent parameters.
Rather, they should lie in specific ranges of permittivity, due to the
simple physical constraint on geometry $a_c/a>1$.

From a practical standpoint, an electrically thin dielectric cylinder,
for which the $c_0^{\rm TM}$ coefficient is dominant, may always be
cloaked by a thin shell with $\epsilon_c<\epsilon_0$.  Under the first
of the conditions in Eq.~\ref{eq:FOT-gen}, most of the scattering may
be suppressed, although significantly negative $\epsilon_c$ may be
required to achieve a very thin shell.  By duality, an electrically
thin magnetic cylinder will require $\mu_c<\mu_0$.

For such infinite cylinders, the cloaking conditions may in general
become trickier than in the spherical case, since the azimuthally
symmetric cylindrical waves follow special cloaking conditions: they
are permittivity-based for zeroth-order when higher order are
permeability-based, and vice versa.  This is particularly relevant for
the conducting scenario, for which the dominant $c_0^{\rm TM}$
coefficient may not be canceled at all in the long-wavelength limit
(cf.  Eq.~(\ref{eq:FOT-PEC})).  It should be emphasized that it is
still possible to suppress scattering in the fully dynamic scenario,
since it may always be canceled with its dynamic expression $U_0^{\rm
  TM}$.  However, in the long-wavelength limit, the required wave
number in the cloak considerably grows, implying that quasi-static
considerations such as those used in Eq.~(\ref{eq:FOT-PEC}) may not be
applied.  However, it is evident that conducting objects are
significantly more challenging to cloak for cylinders than
spheres~\cite{Alu2007a}.  The physical reason for this difference is
evidently related to the fact that electrically thin cylinders are
{\it not} electrically small objects.  These systems are still
infinite in the axial direction (for this analytical treatment), which
implies that their scattering may not necessarily be small -- above
all when conduction currents may be induced in the $z$ direction, as
with the TM$_z^0$ cylindrical wave.

\subsubsection{Validation of the quasi-static conditions: Electrically small dielectric cylinder}

\begin{figure}[!hb]
\centering
\includegraphics[width=16.5cm]{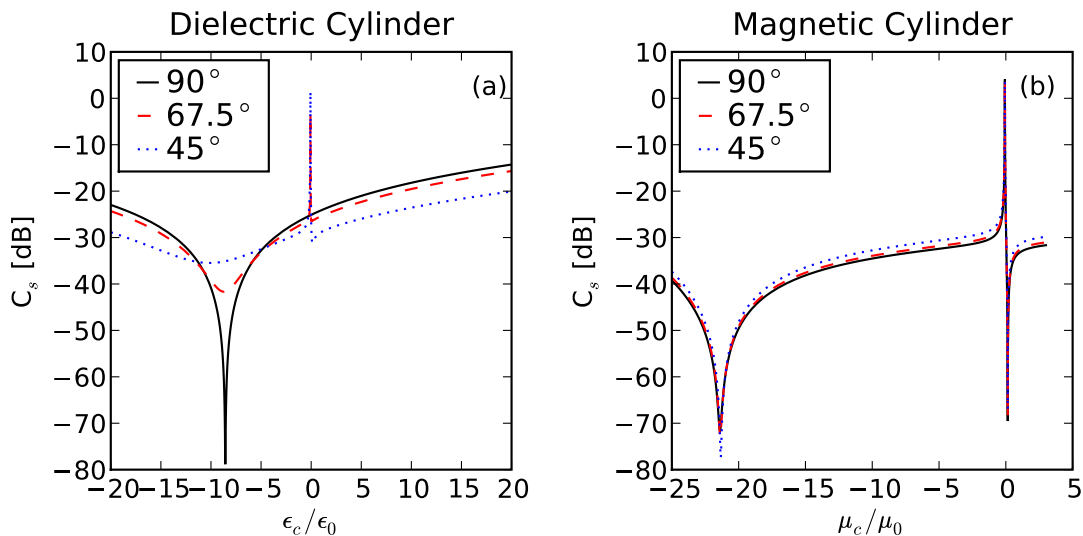}
\vspace*{-1mm}
\caption{Total scattering efficiency for electrically thin
  infinite-length circular cylinders with $k_0a_c=0.1$ and $a_c=1.1a$: \\
  (a) dielectric ($\epsilon=3\epsilon_0$, $\mu=\mu_0$) and (b) magnetic
  ($\epsilon=\epsilon_0$, $\mu=3\mu_0$); for three different
  angles of incidence of TM$_z$-polarized radiation.  In the
  electrically thin limit, the cloaking conditions have weak dependence on
  the angle of incidence.}
\label{fig:th-oblique}
\end{figure}

To assess the validity of the previous approximate cloaking conditions
in the fully dynamic scattering problem, we present some numerical
examples of interest in this section, solved using the exact
analytical theory reported above.  Consider first the case of a
dielectric cylinder with $\epsilon=3\epsilon_0$ and normalized
diameter $k_0 a_c=0.1$, covered by a thin uniform cloaking shell with
$a_c/a=1.1$.  Fig.~\ref{fig:th-oblique}a shows the variation of the
total scattering efficiency $C_s$ as a function of the cloak
permittivity and of the angle of incidence of a TM illuminating plane
wave.  It is immediately apparent that the use of
negative-permittivity thin cloaks may significantly reduce the overall
scattering efficiency of the cylinder, and in particular minimum
scattering is obtained very close to the corresponding quasi-static
cloaking condition for $c_0^{\rm TM}$ in Eq.~(\ref{eq:FOT-gen}), which
would yield $\epsilon_c=-8.524\epsilon_0$.  Moreover, minimum
scattering is achieved at the same value of negative permittivity,
independent of the angle of incidence, consistent with the previous
theoretical results, despite some polarization coupling for oblique
angles.  Overall scattering reduction, compared to no cloak
($\epsilon_c=\epsilon_0$), is especially significant in the normal
incidence case, which provides over 50~dB reduction.  For oblique
incidence, the coupling with TE coefficients affects the cloaking
performance and generates residual scattering at the cloaking
condition, but it is seen that the same (negative) value of cloak
permittivity may provide significant scattering reduction (over 10 dB)
for a wide angular range without drastic changes.

For smaller incidence angles, we also observe excitation of a
plasmonic resonance for slightly negative values of $\epsilon_c$.
This is associated with the condition $V_1^{\rm TE}=0$, obtained here
in the quasi-static limit for $\epsilon_c=-0.07\epsilon_0$, consistent
with findings on quasi-static plasmonic resonances in layered spheres
and cylinders~\cite{Alu2005b}.  It is evident that the resonance is
not excited at normal incidence, but arises as soon as oblique
incidence is considered, due to cross-polarization coupling.  This
explains the sharp peak for small negative values of $\epsilon_c$ in
Fig.~\ref{fig:th-oblique}a.  Analogously, for low positive values of
$\epsilon_c$, a second small dip in the scattering efficiency appears
at $\alpha=\pi/4$, associated with the cloaking condition for
$c_1^{\rm TE}$ = 0 in Eq.~(\ref{eq:FOT-gen}).  Here, this provides the
solution $\epsilon_c=0.14\epsilon_0$.  Once again, this dip does not
appear at normal incidence, due to the lack of coupling between
polarizations for $\alpha=\pi/2$.  To conclude this discussion, it
should be underlined that, in absolute value, scattering is larger for
normal incidence in the uncloaked scenario ($\epsilon_c=\epsilon_0$ in
the figure), as expected, due to the larger $E_z$ component.  On the
other hand, the proper cloak design may maximally suppress scattering
in this worst-case condition, which is an inherent property of this
cloaking approach.

\subsubsection{ Validation of the quasi-static conditions: Electrically small magnetic cylinder }

As a second example, consider the same cylindrical geometry, but with
magnetic properties ($\mu=3\mu_0$), again excited by a TM$_z$ wave.
All other parameters are kept the same.  The cylinder's scattering
efficiency is shown in Fig.~\ref{fig:th-oblique}b.  As anticipated
above, this configuration gives much weaker scattering in the
uncloaked scenario ($\mu_c=\mu_0$), due to its small size and lack of
dielectric contrast with the background.  In fact, the $c_0^{\rm TM}$
coefficient is negligible in this case. Residual scattering is
dominated by the higher-order $c_1^{\rm TM}$ coefficient, which may be
canceled with $\mu_c=-21.2\mu_0$ or $0.14\mu_0$ (note the dual
behavior compared to $c_1^{\rm TE}$ in Fig.~\ref{fig:th-oblique}a).
Two clear scattering dips are visible around these two values, which
also in this scenario do not depend on the angle of incidence,
consistent with Eq.~\ref{eq:FOT-gen}.  The absence of scattering from
higher-order modes and of coupling for the $n=0$ modes ensures that
overall cloak performance is unchanged by variations of the incidence
angle $\alpha$, as evidenced by the different curves in
Fig.~\ref{fig:th-oblique}b.


\subsubsection{Quasi-Static Conditions: Conclusions}

It should be stressed that the above numerical analyses neglect Ohmic
absorption losses in the cloak and/or core materials.  We include
these effects in the following Sections.  Regardless, the plasmonic
cloaking technique has been shown to be inherently robust against
moderate absorption [20], since it is not based on a resonant effect.
This implies that the curves in Fig.~\ref{fig:th-oblique} would remain
practically unchanged near the cloaking regions when moderate losses
are considered.  On the other hand, the large scattering peaks
associated with plasmonic resonances would be significantly dampened
by realistic losses.  This and the previous considerations imply that:
\begin{itemize}
\vspace{-1mm}
\item[(a)] The plasmonic cloaking technique may be successfully
  applied to infinite cylinders.
\vspace{-2mm}
\item[(b)] The quasi-static conditions,
  Eqs.~(\ref{eq:FOT-gen}-\ref{eq:FOT-PEC}), hold to a very good
  approximation for thin cylinders, $k_0a_c\lesssim 0.1$, and ensure
  dramatic scattering reduction in this limit.  \vspace{-2mm}
\item[(c)] Cloaking conditions for electrically thin cylinders are
  very weakly dependent on $\alpha$.
\vspace{-1mm}
\end{itemize}
Thus, cloak designs for electrically thin cylinders are robust against
the angle of incidence for any scattering order $n$ and polarization.
We explore numerically in the following Sections how these properties
are affected by considering thicker geometries and truncation effects.


\section{Plasmonic Cloaking Design and Optimization}
\label{sec:opt}

For a given infinite cylinder of radius $a$ and permittivity
$\epsilon$, a single-layer plasmonic cloaking shell may be optimized
by varying its two design parameters $a_c$ (or $a_c/a$) and
$\epsilon_c$, using the analytical results derived in
Section~\ref{sec:anal}. In general, it is preferable to choose thin
shell thicknesses, since drastically increased cross-sections imply
reduced bandwidths and larger sensitivities to the design
parameters~\cite{Alu2009b}.

This simple cloaking design inherently limits the overall total
thickness of cylinders that we may cloak, since only few scattering
orders may be drastically suppressed independently. Magnetic
properties of the cloak~\cite{Alu2008b} as well as multi-layer
designs~\cite{Alu2008e} may be considered to increase the available
degrees of freedom and size of objects to be cloaked.  In this work,
however, for sake of simplicity we limit our interest to non-magnetic
cloaks, and we consider moderate cross-sections of the objects to be
cloaked.  In particular, in the following we consider cylinders with
diameters $2a=(\frac{\lambda_0}{2},\frac{\lambda_0}{4},
\frac{\lambda_0}{8})$ and relative permittivities of 3 and 10 in the
dielectric scenario, in addition to a PEC cylinder.  Since the
previous theoretical discussion shows that infinite cylinders may be
effectively cloaked, as anticipated, we consider overall lengths L of
several times the diameter, which may be comparable to or larger than
the wavelength of operation.

Because of the TM--TE polarization coupling described in the previous
Section (cf. also the Appendix~\ref{app}), completely general cloak
optimization would be quite complicated, and in general depend on the
incidence angle and polarization of excitation.  We choose instead to
optimize cloak response at normal incidence -- which is usually the
angle for which larger scattering is produced -- then examine the
response at oblique angles with numerical simulations.  More extensive
cloak optimization would involve the choice of a cost function and the
application of global (e.g.\ genetic algorithms~\cite{Kerkhoff2009})
and local (e.g.\ Quasi-Newton) optimization techniques across all
angles and polarizations, also as a function of the observer's
position.  In general, further optimization may be achieved by
considering magnetic cloaks, as discussed above.  We will analyze
these aspects in the near future.


\subsection{Design Parameter Space: the View from Above}
\label{sub:paramspace}

To provide insight into the effects of the various parameters involved
in cloak design, we examine the variation of scattering efficiency
with $a_c/a$ and $\epsilon_c$ for TE or TM plane wave normal incidence
in the infinite (2D) scenario, using the formulation developed in the
previous Section.  To that end we write the scattering gain as:
\bq
\label{eq:CostFn}
Q_s \biggl( \frac{a_c}{a},\epsilon_c \biggr) \; = \;
\frac{C_s}{C_s^0}
\; .
\eq
$C_s$ is the scattering efficiency from Eq.~(\ref{eq:sum}), and the
superscript $0$ represents the uncloaked case, calculated with
Eqs.~(\ref{eq:Mie}-\ref{eq:Vn}), in the limit $a_c\to a$ and
$\epsilon_c\to\epsilon$.

In general, the above discussions show that it is not possible to
achieve large scattering reduction for both TM and TE illumination
simultaneously with a single-layer permittivity cloak at the same
frequency.  For moderate cross-sections, however, such as those
considered here, scattering is generally dominated by one of the two
responses.  It is thus best to design for maximal suppression of that
polarization.  Hence, we focus on TM$_z$ waves, which dominate
scattering from infinite dielectric and conducting cylinders in the
long wavelength limit (cf. quasi-static analysis in the previous
Section).  Ideally, the cloak is designed to provide minimized
scattering gain for normal incidence.

\begin{figure}[!b]
\vspace*{-1mm}
\centering
\includegraphics[width=7.2cm]{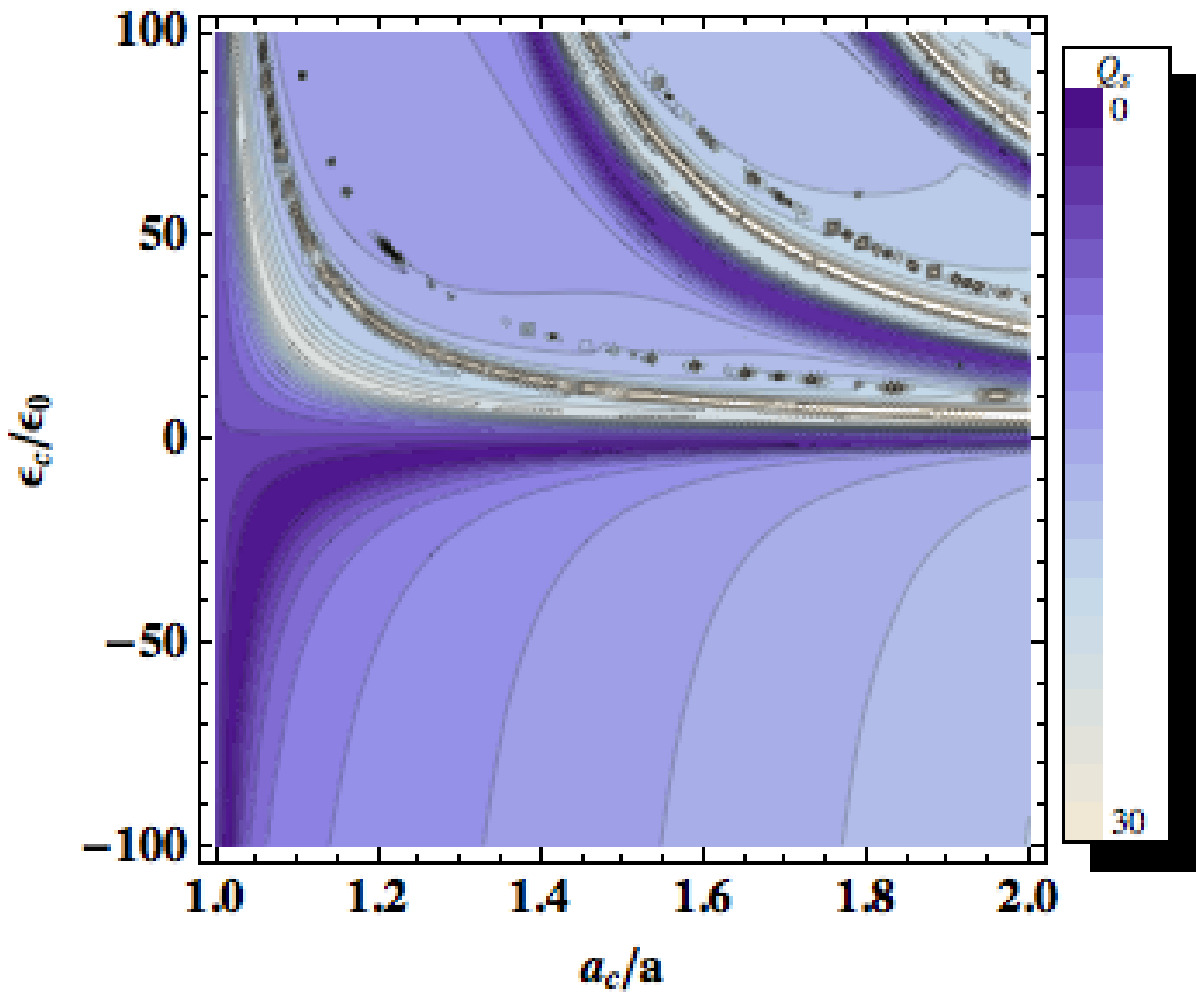}
\includegraphics[width=7.2cm]{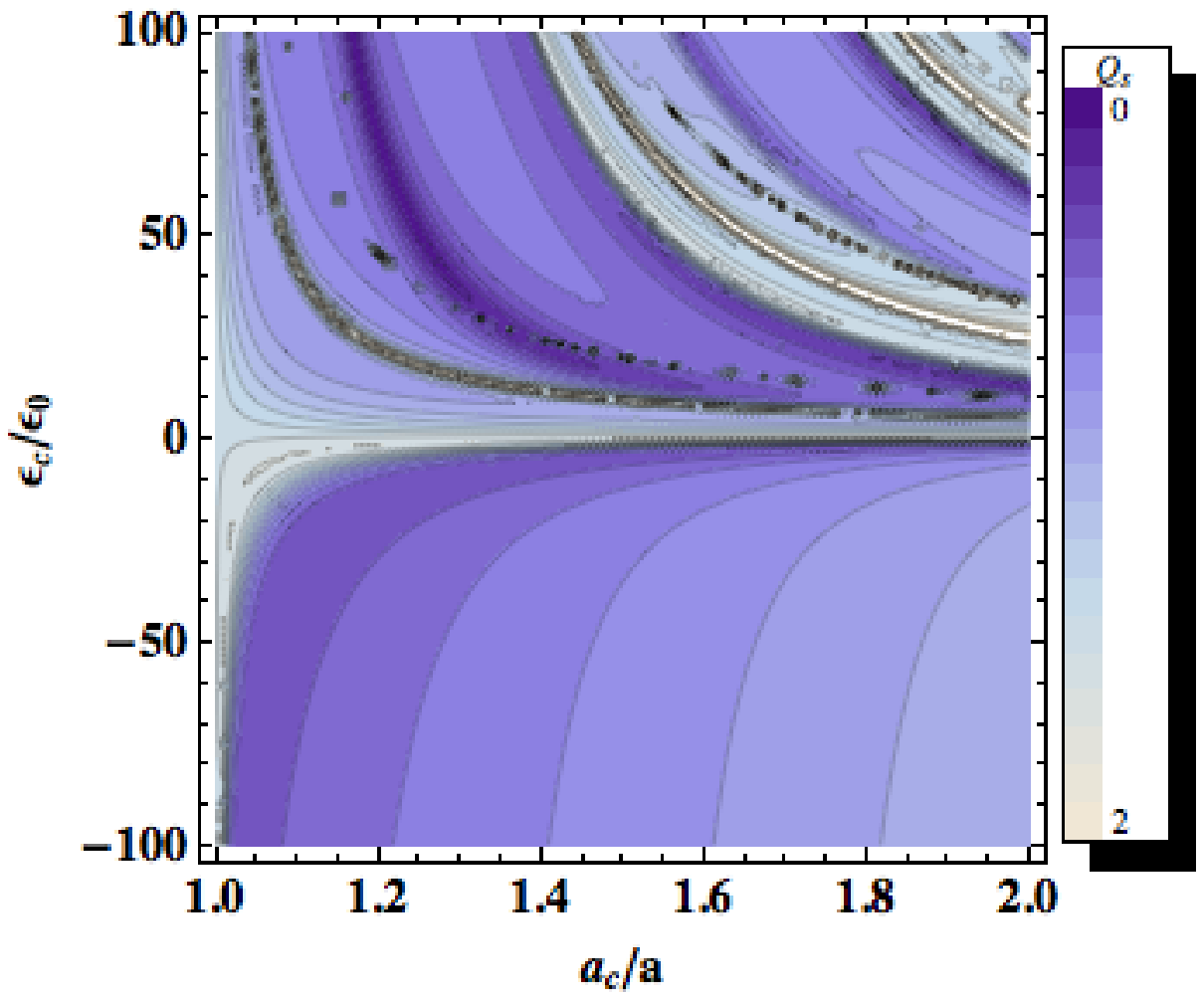}\\
(a) $\epsilon= 3\epsilon_0$
\qquad\qquad\qquad\qquad\qquad\qquad\qquad\qquad
(b) $\epsilon=10\epsilon_0$
\vspace*{-2mm}
\caption{Scattering gain contour plots over cloak parameter space,
  $\epsilon_c$ ($y$-axis) and $a_c/a$ ($x$-axis), for core
  permittivities as labeled.  Note the resonant enhancements and the
  cloaking regions.}
\label{fig:contourscan}
\end{figure}

\begin{figure}[!t]
\centering
\includegraphics[width=7.2cm]{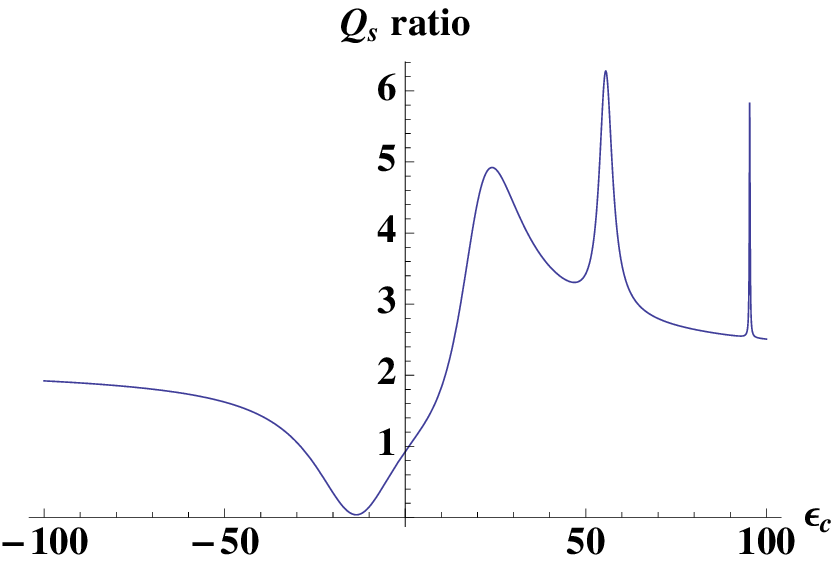}
\hspace{5mm}
\includegraphics[width=7.2cm]{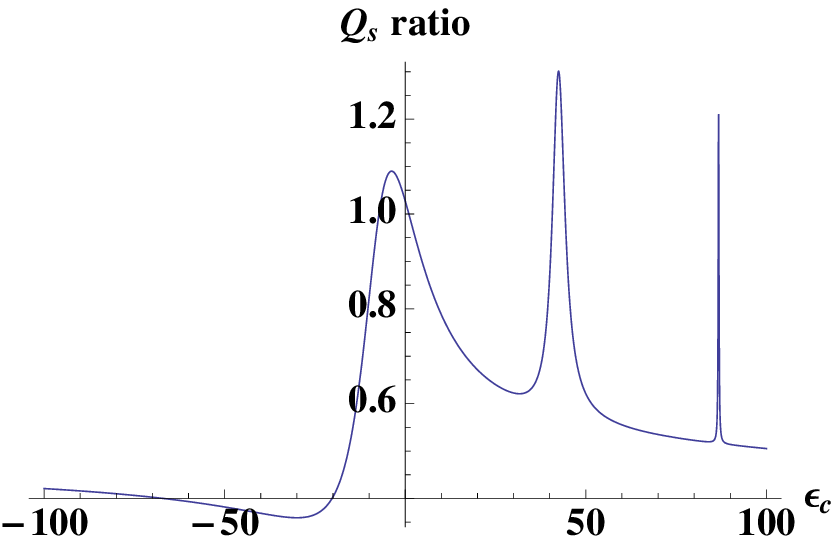}
\caption{The ratio $Q_s$ as a function of cloak permittivity,
  $\epsilon_2$ for slices of constant cloak thickness, $a_c=1.1a$, in
  the contour plots of Fig.~\protect\ref{fig:contourscan}, for core
  permittivities $\epsilon=3$ (left) and $\epsilon=10$ (right).}
\label{fig:eps-opt}
\vspace*{-3mm}
\end{figure}

For our calculations we used Mathematica~\cite{Mathematica} and
truncated the Mie summations at order $n=5$, which ensured convergence
for the cross-sections considered here.  Fig.~\ref{fig:contourscan}
shows scattering gain contour plots for infinite cylinders of diameter
$2a=\lambda_0/4$, relative permittivities $\epsilon=3$
(Fig.~\ref{fig:contourscan}a) and $\epsilon=10$
(Fig.~\ref{fig:contourscan}b), and TM normal incidence.  The plots
show alternating loci of resonant peaks (large $Q_s$, light color) and
cloaking regions (near-zero $Q_s$, dark color).  As expected, thinner
cloaks (smaller $a_c/a$) yield larger bandwidths, reflected by wider
cloaking regions.  This implies that the ideal cloak design would
utilize $\epsilon_c<0$, as expected from the results in
Sec.~\ref{sec:anal}, from simple physical considerations and analogous
results in the spherical scenario~\cite{Alu2005a}.  The cloaking loci
for large positive values of $\epsilon_c$ are associated with
anti-resonances that arise when the shell thickness is comparable to
the wavelength in the shell material.  These anti-resonances are
narrow-bandwidth and highly sensitive to the design parameters, and
they necessarily lie in close proximity with scattering resonant
peaks, which makes them not suitable for a robust cloaking design.
Fig.~\ref{fig:eps-opt} shows constant $a_c/a$ slices of
Fig.~\ref{fig:contourscan} to better illustrate these characteristics.
These slices illustrate how $\epsilon_c<0$ choices typically produce
lower scattering gain, i.e.  better cloaking, away from dangerous
resonant enhancements.  This is especially true when cloaking
low-density dielectric cylinders.

%
%

%

We also considered optimization of plasmonic cloaks for PEC cylinders
of the same size.  However, as noted in the quasi-static limit, here
the dominant scattering contribution (via the $c_0^{\rm TM}$
coefficient), may not be canceled in the long-wavelength regime.
This implies that in the corresponding contour plots there would not
be cloaking regions for thin negative-permittivity shells, as in the
dielectric scenario.  Moderate scattering reduction may be achieved
with large-$\epsilon_c$ thick shells, but as outlined above this
implies large sensitivity to frequency, design variations and
closely-spaced resonant peaks, which may be excited at different
incidence angles.  It is thus evident that a simple permittivity cloak
may not be sufficient to adequately cloak a PEC cylinder and, as in
the spherical scenario~\cite{Alu2005a}, it may require magnetic
permeability $\mu_c$ different from the background for robust
scattering reduction.  In the following, therefore, we mainly focus
our design efforts on dielectric cylinders.


\subsection{Cloak Design}
\label{sub:opt}

The previous results show that optimal cloak configurations for
dielectric cylinders are based on negative permittivity metamaterials
and thin cloak shells. We should point out that negative values of
effective permittivity may indeed be achieved in the microwave or THz
frequency ranges using various metamaterial geometries, such as wire
media or parallel-plate implants~\cite{Silveirinha2007}, and they are
naturally available at larger frequencies.  In particular, the
parallel-plate implant technology may be particularly well-suited for
cloaking incident TM$_z$ waves, as it has been already demonstrated
theoretically and verified experimentally for normal incidence at
microwave frequencies~\cite{Silveirinha2007,Edwards2009}.  We assume
here, for sake of simplicity, that the required value of effective
permittivity is available for the shell geometry of interest and that
the cloak material is isotropic.  Possible anisotropy for a specific
metamaterial realization, inherent in some proposed
realizations~\cite{Silveirinha2007,Edwards2009}, may affect cloak
performance for different incidence angles, but our preliminary
results show that for thin cloaks such effects may be minor.  Thus,
for simplicity we always assume idealized isotropic metamaterials in
the following.  The requirement that any passive metamaterial has a
frequency dispersion ensuring
$\partial(\omega\epsilon_c)/\partial\omega>0$ is met here by assuming
a Drude dispersion model of the form:
\bq
\label{eq:Drude}
\epsilon_c(\omega)
\, = \,
 1 - \frac{\omega_p^2}{\omega(\omega-j\gamma)}
\, .
\eq
Here and in the numerical simulations in the next Section, we
calculated the plasma frequency $\omega_p$ to ensure that at the
design frequency $f_0$ the real part of $\epsilon_c$ yields the
required value.  The damping frequency $\gamma$, associated with the
level of losses in the metamaterial, has been always assumed in the
following to be $\gamma=10^{-2}\omega_p$.  This value provides a
moderate amount of loss, comparable with practically-realizable
metamaterial geometries at these frequencies.

Table~\ref{tab:opt} summarizes an extensive campaign of optimizations
that we performed to cloak different cylindrical geometries, as
outlined above, considering several cloaking designs.  Inspecting
Table I, we see that thin cylinders ($2a=\lambda/8$), consistent with
the previous quasi-static analysis, require optimized cloaks with
negative permittivity in the case of dielectric objects, and a large
positive permittivity for conducting materials.  The corresponding
scattering gain may become extremely low, since only few Mie
coefficients effectively contribute to the total scattering, and they
are properly canceled by the optimized cloak design.  As a general
rule of thumb, consistent with the previous considerations, thinner
cloaks require larger values of permittivity, either positive or
negative.  Relatively thicker cloaks relax the requirements on very
negative permittivity of the cloak, but, as highlighted above, larger
thickness also tends to produce additional scattering terms, which
limits overall performance.  However, lower absolute values of
negative (in the case of dielectric) or positive (for PEC)
permittivity may be easier to achieve and be less sensitive to loss.
A trade-off between cloak thickness, material reliability and overall
cloaking performance should be considered.  Larger dielectric objects
may be cloaked by thicker cloaks with positive values of permittivity,
since the dynamic nature of the wave in the cloak may produce a
negative polarizability even with positive materials in larger shells.
However, the overall scattering reduction is less dramatic; residual
scattering is indeed expected for larger objects.  In particular, as
discussed above, conducting cylinders are most challenging to be
cloaked against TM waves, due to the axial conduction currents induced
on their surface.  Scattering reduction of around 3~dB, however, may
still be achieved even for conducting cylinders.  Moreover, large
values of core permittivity $\epsilon$ are harder to cloak, and the
residual scattering is larger than for a lower-permittivity core of
the same size.

Our cloak designs in Table~\ref{tab:opt} were optimized for TM normal
incidence, the scenario with maximum scattering for the considered
objects.  At oblique incidence, however, TM--TE coupling excites
additional scattering modes, which may affect the overall response, in
particular for thicker objects and moving toward grazing incidence.
We discuss these features in detail in the next Section.

\begin{table}[!t]
\centering
\begin{tabular}{|c|c|c|c|c|}
\hline
\, Core diam. ($2a$) \, & \, Core $\frac{\epsilon}{\epsilon_0}$ \,
                        & \, Cloak $\frac{a_c}{a}$ \,
                        & \, Cloak $\epsilon_c$ \,
                        & $Q_s$ \\
\hline\hline
\multirow{5}{*}{$\frac{\lambda}{2}$}
  &   3 & 1.10 & -8.16 & 0.26 \\[-1mm]
  &   3 & 1.40 & 22.45 & 0.13 \\[-1mm]
  &  10 & 1.05 & 13.37 & 0.22 \\[-1mm]
  &  10 & 1.10 &  6.91 & 0.22 \\[-1mm]
  & PEC & 1.10 & 95.48 & 0.47 \\
\hline
\multirow{5}{*}{$\frac{\lambda}{4}$}
  &   3 & 1.05 & -27.88 & 0.031 \\[-1mm]
  &   3 & 1.10 & -13.55 & 0.038 \\[-1mm]
  &  10 & 1.10 & -35.00 & 0.36  \\[-1mm]
  &  10 & 1.20 &  74.57 & 0.16  \\[-1mm]
  & PEC & 1.50 &  14.01 & 0.37  \\
\hline
\multirow{4}{*}{$\frac{\lambda}{8}$}
  &   3 & 1.05 & -20.26 & 0.00076 \\[-1mm]
  &   3 & 1.10 &  -9.45 & 0.00092 \\[-1mm]
  &  10 & 1.10 & -56.25 & 0.0017  \\[-1mm]
  &  10 & 1.30 & -17.87 & 0.0034  \\[-1mm]
  & PEC & 1.40 &  88.92 & 0.096   \\
\hline
\end{tabular}
\caption{Parameters from the cloak optimization procedure described
  in Sec.~\protect\ref{sub:opt} for different objects and cloak
  thicknesses.  $Q_{s}$ corresponds to the scattering gain for
  normal-incidence TM$_z$ illumination.}
\label{tab:opt}
\end{table}

\begin{figure}[!hb]
\centering
\includegraphics[width=17cm]{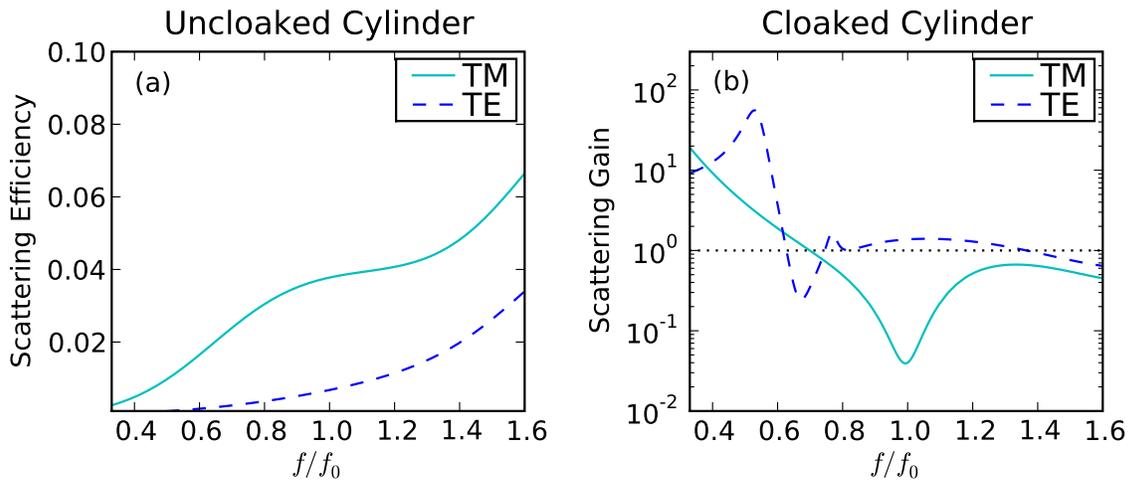}
\vspace*{-3mm}
\caption{Uncloaked scattering efficiency (a) and expected scattering
  gain (b), $C_s/C_s^0$, as a function of relative frequency, for an
  infinite cylinder with $2a=\lambda_0/4$, $\epsilon=3$ and for
  $a_c=1.1a$ (see Table~\protect\ref{tab:opt}).  Optimized for TM
  waves at normal incidence, the cloak will perform differently for TE
  illumination.}
\label{fig:TMvTE}
\end{figure}

Fig.~\ref{fig:TMvTE} considers the typical frequency dependence of
cloak performance on wave polarization.  The left panel shows the
uncloaked scattering widths for an infinite dielectric cylinder with
$2a=\lambda_0/4$ and $\epsilon=3$ for TM and TE normally-incident
plane waves.  As expected, the TM polarization is more strongly
scattered, thus it is of more relevance in designing an optimized
cloak.  However, despite the much smaller response to TE waves, it is
interesting to analyze the behavior of a TM-optimized cloak to TE
polarization, since a portion of the TM scattered power may be coupled
to TE waves for TM oblique incidence, in particular for smaller
incidence angles.  The right panel shows the normalized scattering
gain for TM and TE incidence, considering a conformal shell cloak with
$a_c=1.1a$ and $\epsilon_c=-13.55$ at design frequency $f_0$ with
Drude dispersion, which coincides with the optimized design at $f_0$
for TM normal incidence.  We see how the cloak is very effective over
a relatively broad range of frequencies around $f_0$ for TM incidence,
despite the natural dispersion of the cloak material.  However, this
same cloak does not provide scattering reduction at the same frequency
for impinging TE waves, supporting instead a small cloaking dip at
lower frequencies. Although this {\it per se} is not a major issue,
since the absolute value of the TE scattering is much smaller (left
panel), and the quantities in the right panel are normalized to the
corresponding uncloaked scattering, this issue may cause an overall
dependence of the cloaking response to the incidence angle for TM
excitation, due to the TM-TE polarization coupling.  We verify these
issues in the next Section, which considers the performance of
finite-length cylinders illuminated at oblique incidence.

%


\newpage

\section{Numerical Results for Finite-Length Cylinders and Oblique Incidence}
\label{sec:num}

After having considered the optimization of cloaks for dielectric and
conducting infinite cylinders, we consider in this Section the effects
of finite length and varying angle of incidence on the performance of
cylindrical plasmonic cloaks, using full-wave numerical
simulations~\cite{CST}.  To keep the design simple, in our numerical
simulations we simply truncated the cloaked cylinder, as optimized in
the previous section, consistent with the geometry in
Fig.~\ref{fig:cyl}, leaving the ends uncovered.  This is expected to
cause additional scattering, especially for more oblique incidence
(smaller $\alpha$).  However, since this cloaking technique is based
on an integral effect~\cite{Alu2007a}, for moderate cross-sections
these effects are not expected to significantly deteriorate the
overall cloaking performance.  This intuition is confirmed by the
following numerical results.  Improved performance may be obtained by
locally tailoring the cloak around the edges, and covering the ends,
as described for conical geometries in~\cite{Tricarico2010}.  The
small improvements achievable with this fine-tuning of the cloak
design are in any case not relevant to the following general
discussion.

\subsection{Dielectric Cylinder at Normal Incidence}
\label{sub:primary}

Consider first the dielectric cylinder of Fig.~\ref{fig:th-oblique}a
($\epsilon=3\epsilon_0$, $\mu=\mu_0$, diameter $2a=2.5$~cm), but with
finite length $L=10$~cm.  Our objective is to reduce its scattering
cross-section at the design frequency $f_0=3$~GHz, for which this
object has cross-sectional width $2a=\lambda_0/4$ and length
$L=\lambda_0$; $\lambda_0=10$~cm is the free-space wavelength at
frequency $f_0$.  In this frequency range, the object may produce
significant scattering, since its length and electrical size are
comparable to the incident wavelength.  Scattering at normal incidence
is dominated by the TM$_z$ polarization (cf.  Sec.~\ref{sec:anal}), so
we excite the cylinder with an impinging TM wave, and use a cloak with
permittivity $\epsilon=-13.55\epsilon_0$ ($\omega_p=11.44$~GHz) at
$f_0$ and thickness $a_c=1.1a$, obtained in Sec.~\ref{sec:opt}.  Note
the relatively large negative permittivity required by such thin
layer.  For a thicker cloak, $|\epsilon_c|$ may be reduced, but at the
price of possible excitation of higher-order cylindrical harmonics,
which may slightly deteriorate the cloak performance.

\begin{figure}[!b]
\centering
\vspace*{-5mm}
\includegraphics[width=16cm]{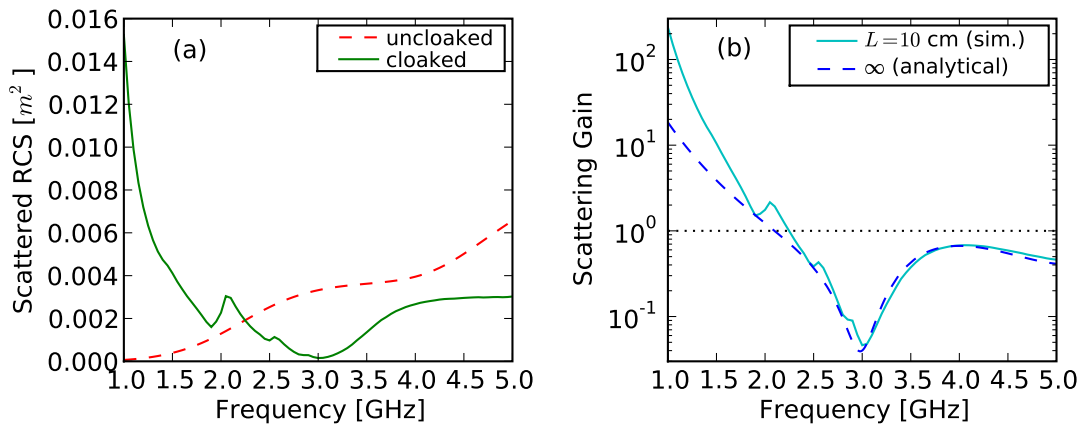}
\vspace*{-1mm}
\caption{(a) Frequency dispersion of the total RCS for a dielectric
  cylinder with $\epsilon=3\epsilon_0$, $\mu=\mu_0$, $2a=2.5$~cm and
  finite length $L=10$~cm, comparing uncloaked v. cloaked (as
  optimized in Section~\protect\ref{sec:opt}) with $a_c=1.1a$ and
  Re[$\epsilon_c$]$=-13.55\epsilon_0$ using a Drude model at
  $f_0=3$~GHz for TM illumination at normal incidence
  ($\omega_p=11.44$~GHz).  (b) The corresponding scattering gain
  (ratio of cloaked to uncloaked RCS), compared with analytical
  results obtained for an infinite cylinder with otherwise identical
  geometry.  The full-wave simulations were performed
  with~\protect\cite{CST}.}
\label{fig:UvC}
\end{figure}

Fig.~\ref{fig:UvC}a plots the total 3D radar cross section
(RCS)~\cite{Balanis1989}, obtained by integrating the scattering cross
section over all angles, calculated using commercial software based on
the finite-integration technique ~\cite{CST}.  It compares the cloaked
(solid green curve) and uncloaked (red dashed curve) scenarios for
normal incidence ($\alpha=\pi/2$).  Despite the cylinder's finite
electrical length and its non-negligible cross-section, scattering at
the design frequency may be significantly suppressed with a proper
cloak design.  As expected, the scattering is a minimum at $f_0$, and
the bandwidth of operation is moderately large, as is common for
plasmonic cloaks~\cite{Alu2007a}.  Significant scattering reduction is
achieved over a fractional bandwidth of over $30\%$.
Fig.~\ref{fig:UvC}b plots the corresponding scattering gain, defined
as the ratio between the total RCS of the cloaked and uncloaked
geometries.  The curves effectively show the overall scattering
reduction achieved by adding the cloak to the bare cylinder for
different frequencies.  The plot compares the full-wave simulations
for finite-length cylinders calculated using~\cite{CST}, with the
infinite-cylinder analytical formulation derived in the previous
section.  The overall scattering reduction is $\approx 15$~dB at
$f_0$, and the analytical result for an infinite cylinder agrees
extremely well with the full-wave simulation for a finite cylinder,
despite the truncation effects.  At lower frequencies, $\lesssim
2$~GHz, the plasmonic nature of the cloak and its negative
permittivity induces a plasmonic wave along the cylinder, which
actually increases the overall scattering compared to the uncloaked
scenario.  This is predicted in both the infinite and finite-length
cases, and readily observed in Fig.~\ref{fig:UvC}b.  There, the
contribution is even stronger for the finite-length truncated
cylinder, due to plasmonic radiation at the ends.

Table~\ref{tab:num} summarizes the calculated overall suppression at
$f_0$ for the dielectric-core $a_c=1.1a$ cloak designs of
Table~\ref{tab:opt} illuminated by TM waves at normal incidence.  In
particular, the table compares the simulated scattering gain for
finite lengths $C_{s,sym}$ with the theoretical value predicted
analytically for infinite cylinders $C_{s,th}$.  We observe excellent
agreement between predicted results from the analytical formulas for
the infinite cylinder, and full-wave numerical simulations for the
truncated geometries.  This implies that truncation effects are not
significant in several realistic setups when cloaking dielectric
cylinders with the plasmonic cloaking technique.

\begin{table}[!t]
\centering
\begin{tabular}{|c|c|c|c|c|c|}
\hline
\, Core diam. ($2a$) \, & \, $L/2a$ \,
                        & \, Core $\frac{\epsilon}{\epsilon_0}$ \,
                        & \, Cloak $\epsilon_c$ \,
                        & $Q_{s,th}^{\rm TM}$
                        & $Q_{s,sim}^{\rm TM}$ \\
\hline\hline
$\frac{\lambda}{2}$
  & 2  &   3 & -8.16 & 0.17 & 0.27 \\
\hline
\multirow{2}{*}{$\frac{\lambda}{4}$}
  & 4  &   3 & -13.55 & 0.038 & 0.046 \\[-1mm]
  & 4  &  10 & -35.00 & 0.17  & 0.38  \\
\hline
$\frac{\lambda}{8}$
  & 4 &   3 &  -9.45 & 0.0049 & 0.042 \\
\hline
\end{tabular}
\caption{Scattering gain at frequency $f_0$ for the cloak designs of
  Table~\protect\ref{tab:opt} with $10\%$ thickness ($a_c=1.1a$), and
  TM$_z$ normal incidence.  We compare the theoretical results for
  infinite cylinders ($Q_{s,th}$) with the numerical full-wave
  simulations for truncated cylinders with $L=\lambda_0$ ($Q_{s,sim}$).}
\label{tab:num}
\vspace*{-3mm}
\end{table}
\subsection{Varying Angle of Incidence}

\begin{figure}[!b]
\vspace*{-5mm}
\includegraphics[width=17cm]{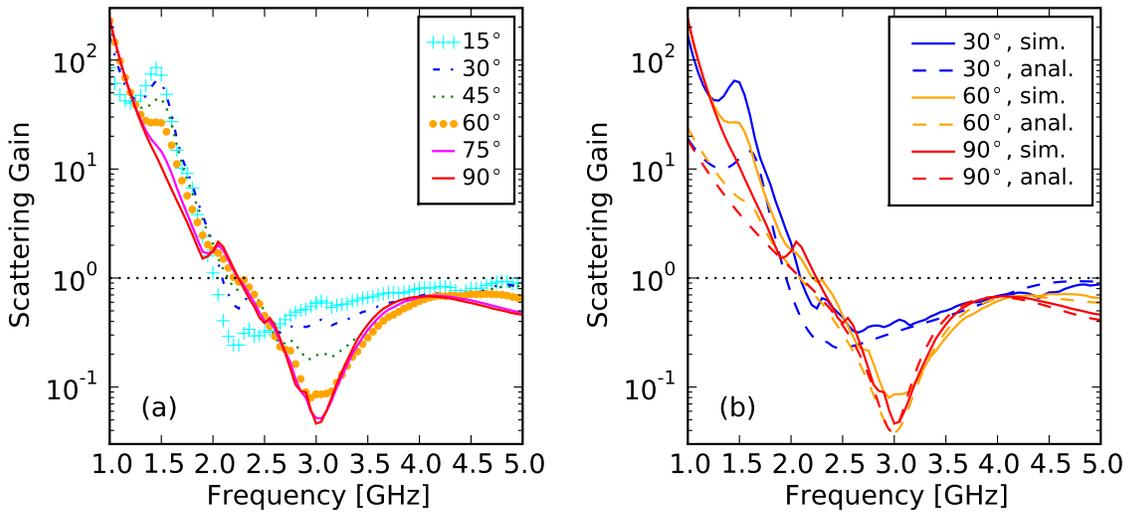}
\vspace*{-1mm}
\caption{Scattering gain for the cylinder of
  Fig.~\protect\ref{fig:UvC}, for various angles of incidence
  $\alpha$.  (a) Results from numerical simulation.  (b) Numerical
  simulation (solid) v.\ analytical infinite-cylinder results (dashed)
  for select angles.}
\label{fig:angles}
\end{figure}

Fig.~\ref{fig:angles} shows the variation of scattering gain for the
cylinder of Fig.~\ref{fig:UvC} varying the incidence angle $\alpha$,
as depicted in Fig.~\ref{fig:cyl}.  Even for small (near grazing)
incidence angles, the cloaking effects are barely different from the
normal-incidence case.  For smaller values of $\alpha$, cloaking is
somewhat reduced at $f_0$, due to the excitation of TE scattered waves
via cross-polarization coupling, as discussed in Sec.~\ref{sec:anal}.
These results demonstrate strong agreement between the analytical
calculations for infinite cylinders and the numerical simulations for
finite $L$, although some expected minor deviation does appear at
small angles, due to end effects.  We emphasized that, although the
normalized scattering gain is reduced less for smaller angles, the
overall RCS is significantly smaller in these cases, since the
electric field component along the cylinder is comparatively shorter,
consistent with the discussion in the previous Section.  The results
of Fig.~\ref{fig:angles} demonstrate convincingly that plasmonic
cloaking may be applied to finite-length dielectric cylinders at
oblique excitation, although at small $\alpha$ one may be required to
cloak the TE contribution separately for improved performance. This is
consistent with our findings in Fig.~\ref{fig:TMvTE}.

\subsection{Varying Diameter}

\begin{figure}[!t]
\centering
\includegraphics[width=7cm]{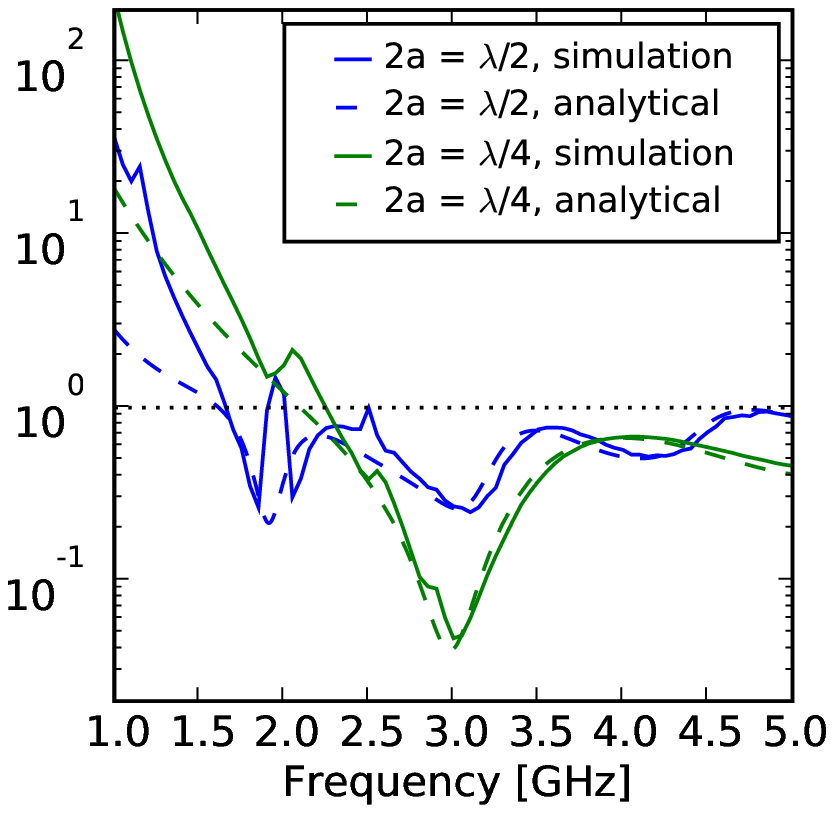}
\hspace{5mm}
\includegraphics[width=7cm]{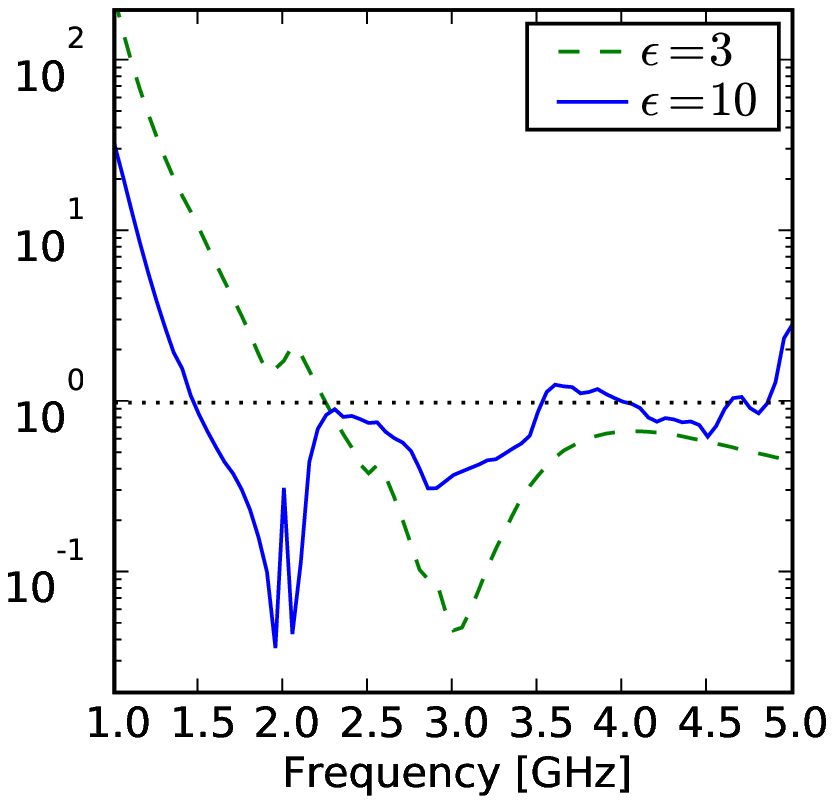}
\vspace*{-1mm}
\caption{Left: Comparison of scattering gain for two diameters of
  dielectric cylinders, as labeled, illuminated at normal incidence
  with a TM polarized wave.  Right: Comparison of scattering gain
  under normal-incidence TM illumination for a cylinder with same
  geometry as in Fig.~\protect\ref{fig:UvC}, but with core
  permittivities $\epsilon/\epsilon_0=3,10$ as labeled.  The
  calculations used the optimized cloak designs as described in
  Table~\protect\ref{tab:opt}.}
\label{fig:diams+eps}
\end{figure}

The left panel of Fig.~\ref{fig:diams+eps} shows the scattering gain
under normal-incidence TM$_z$ illumination for the cylinder of
Sec.~\ref{sub:primary} (cf.  Fig.~\ref{fig:UvC}) next to that for a
cylinder with twice that diameter, $2a=\lambda_0/2$ (cf.
Table~\ref{tab:opt} for cloak parameters).  The figure also compares
these results with the infinite-cylinder analytical result of
Sec.~\ref{sec:anal}.  End effects are negligible except at low $f$,
and the analytical results match very well the numerical simulations
-- especially at frequencies where plasmonic resonances are not
excited.  Cloaking is effective over a relatively broad bandwidth,
even for the thicker cylinder.  This implies that the designed cloaks
may be effective over a relatively broad range of object size.  The
excellent agreement between the analytical curves for infinite
cylinders and the simulation results for truncated geometries indicate
that truncation effects are negligible for cloaking performance and
suggest the use of the previously derived analytical formulas for fast
cloak optimization.  A simple permittivity cloak, as considered here,
is effective for significant scattering reduction for cylinders with
diameters of the order of the wavelength.  Thicker cylinders may
require the use of multilayer and/or magnetic cloaks, as discussed in
Ref.~\cite{Alu2008b} for spherical geometries.

\subsection{Varying Object Permittivity}

The right panel of Fig.~\ref{fig:diams+eps} shows the scattering gain
for the cylinder of Sec.~\ref{sub:primary} (cf. Fig.~\ref{fig:UvC}),
for the same dielectric and also with a denser core,
$\epsilon=10\epsilon_0$ (see Table~\ref{tab:opt}).  In the
large-$\epsilon$ limit, this comparison underscores some of the
challenges that may be involved in cloaking a conducting object, which
would coincide with a dielectric core in the limit of very large
permittivity.  There is significant scattering reduction around the
design frequency $f_0$ also in the denser scenario, but this example
simultaneously supports even stronger cloaking at a lower frequencies,
$f\sim 2$~GHz.  This effect, predicted by the analytical results for
infinite cylinders, is associated with the frequency dispersion of the
cloak, which matches the cloaking condition for the coefficient at
lower frequencies (for more negative values).  Effectively, this
scenario presents a coincidental cloaking effect at another frequency.
This does {\it not} imply that one could tune the second suppression
arbitrarily, since its position depends on the natural metamaterial
dispersion.  However, one could tune its separation from $f_0$ to a
limited degree if the cloak thickness were not fixed at $a_c=1.1a$.

\subsection{Varying Cylinder Length}

\begin{figure}[!pt]
\centering
\includegraphics[width=14cm]{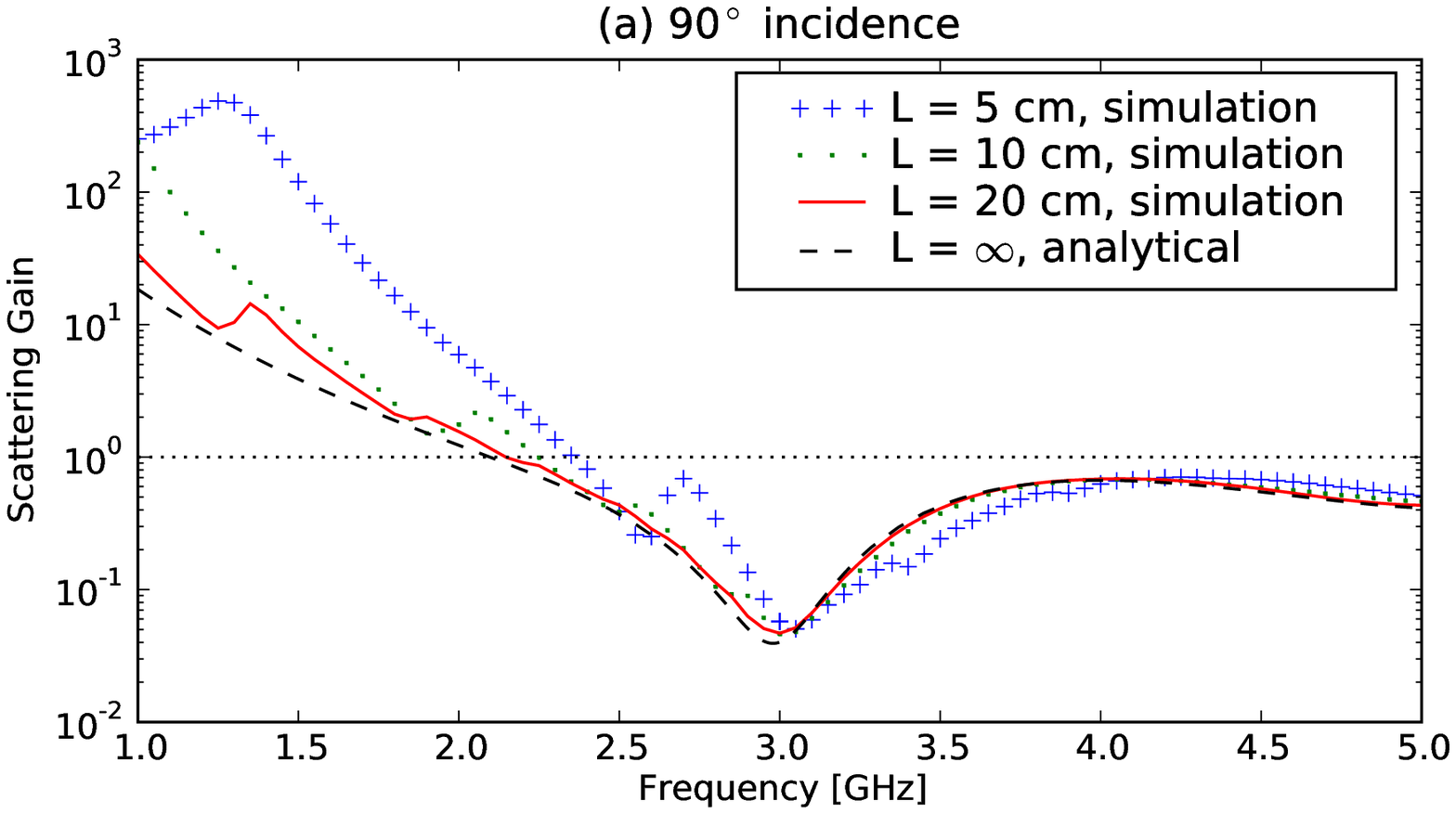}
\includegraphics[width=14cm]{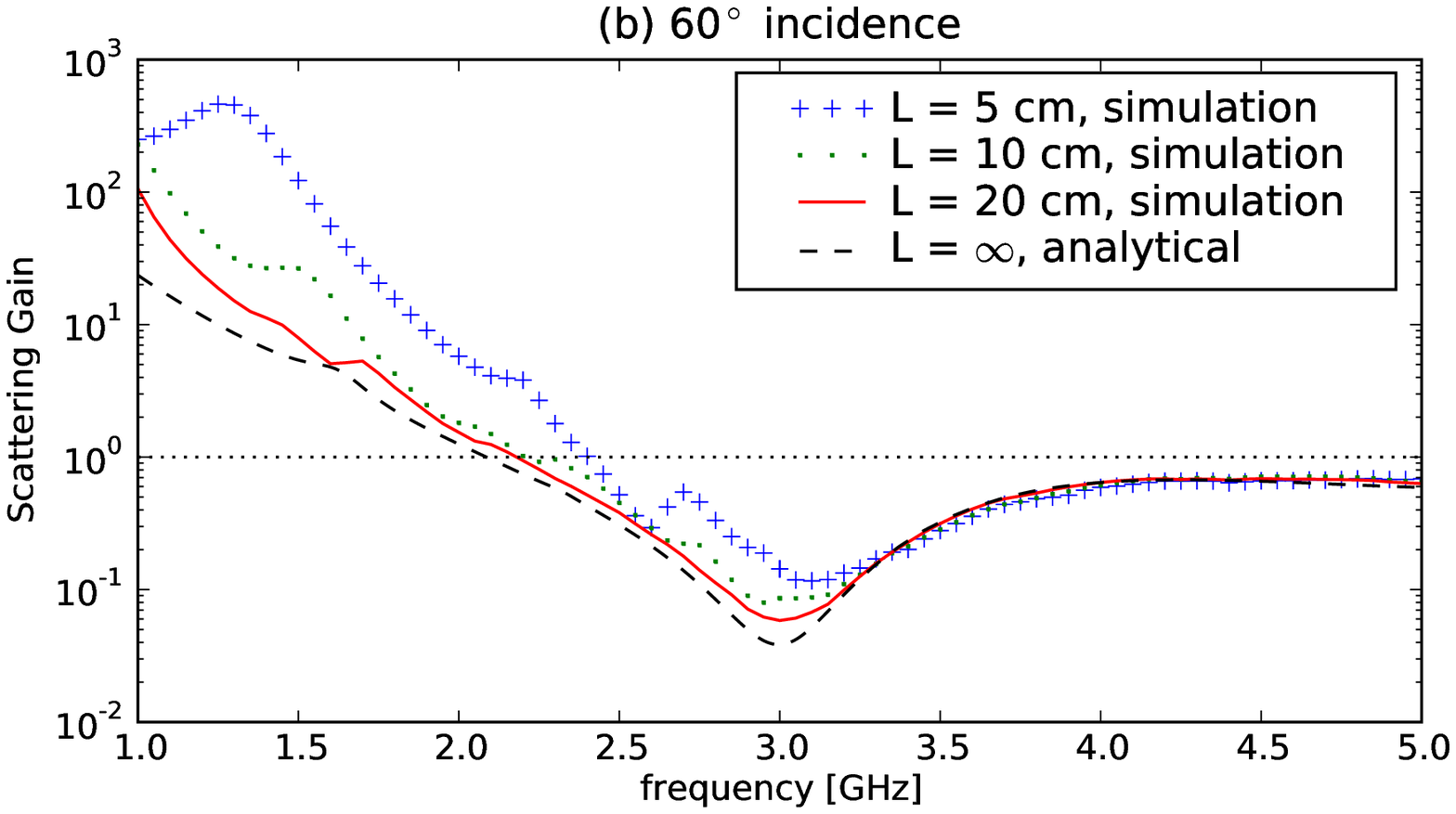}
\includegraphics[width=14cm]{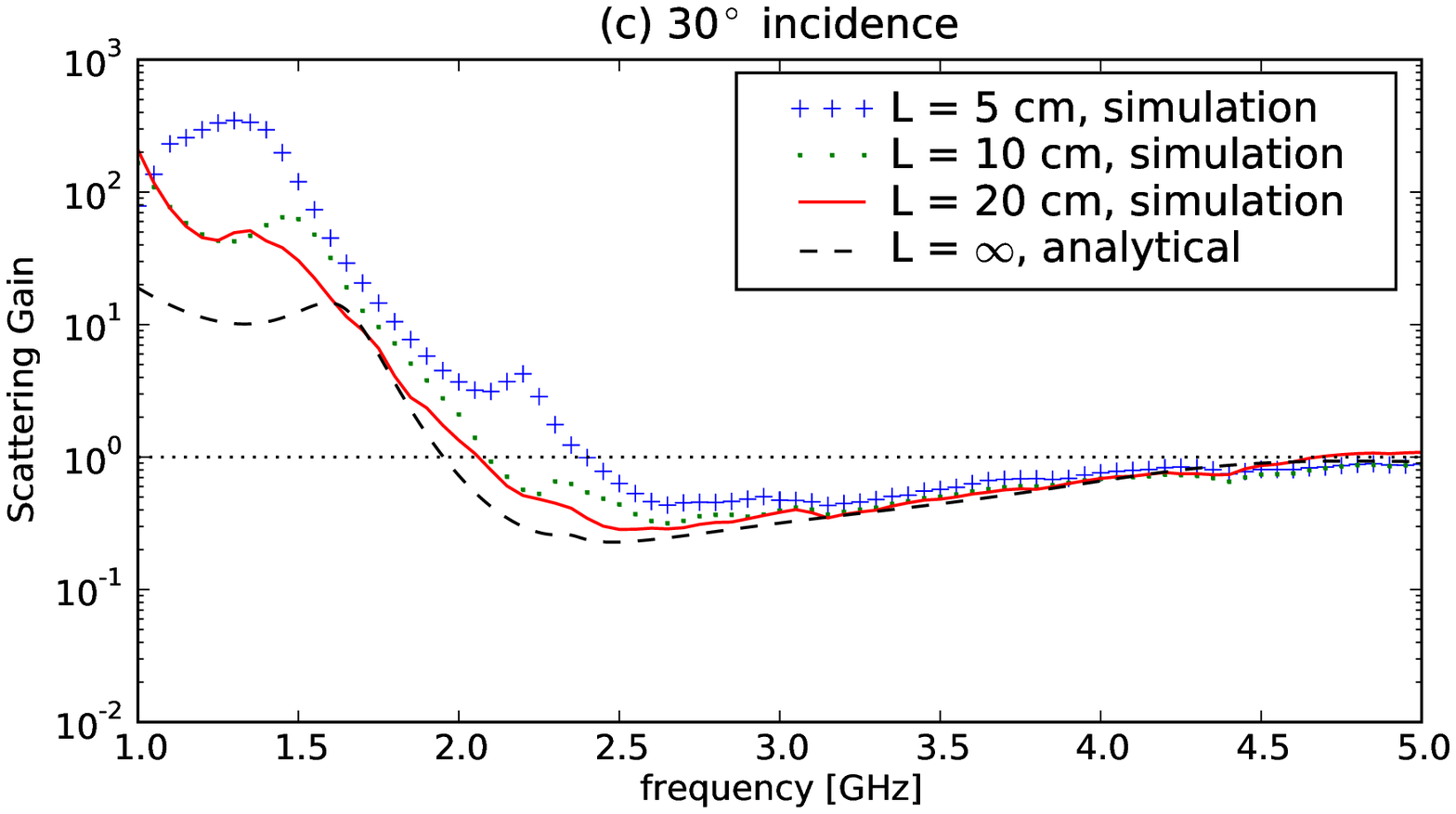}
\caption{Scattering gain for the cylinder of
  Fig.~\protect\ref{fig:UvC} and variations in length by a factor of
  two, for three different angles of incidence under TM excitation.
  The cloaking design is robust even for very short lengths, almost up
  to $L\sim 2a$.}
\label{fig:incidence}
\end{figure}

Fig.~\ref{fig:incidence} analyzes in more detail the effects of
truncation for the case of Fig.~\ref{fig:UvC} in
Sec.~\ref{sub:primary}, by considering different truncation lengths
for several incidence angles: $\alpha=\pi/2,\pi/3,\pi/6$ for the three
panels, respectively.  Each compares different lengths, including the
analytical result for infinite cylinders.  The cloaking effect is
indeed robust against variations in incidence angle, and truncation
effects are quite moderate, as the scattering gain follows the line
calculated analytically in the infinite cylinder geometry.  Even for
the shortest length ($L=5$~cm, a 2:1 aspect ratio), the full-wave
simulation still follows the infinite-length case with surprising
accuracy, above all near $f_0$.  In this case, scattering is
characterized by small resonant peaks at lower frequencies, associated
with longitudinal resonances due to the finite length. The results are
nevertheless extremely encouraging, in particular for shorter
cylinders, for which the cloaking design works extremely well in the
frequency range of interest.

\subsection{Near- and Far-Field Plots}

\begin{figure}[!pt]
\hspace*{-15mm}
(a) uncloaked cylinder
\qquad\qquad\qquad\qquad\qquad\qquad\qquad
(b) cloaked cylinder \\[2mm]
\centering
\includegraphics[width=7.0cm]{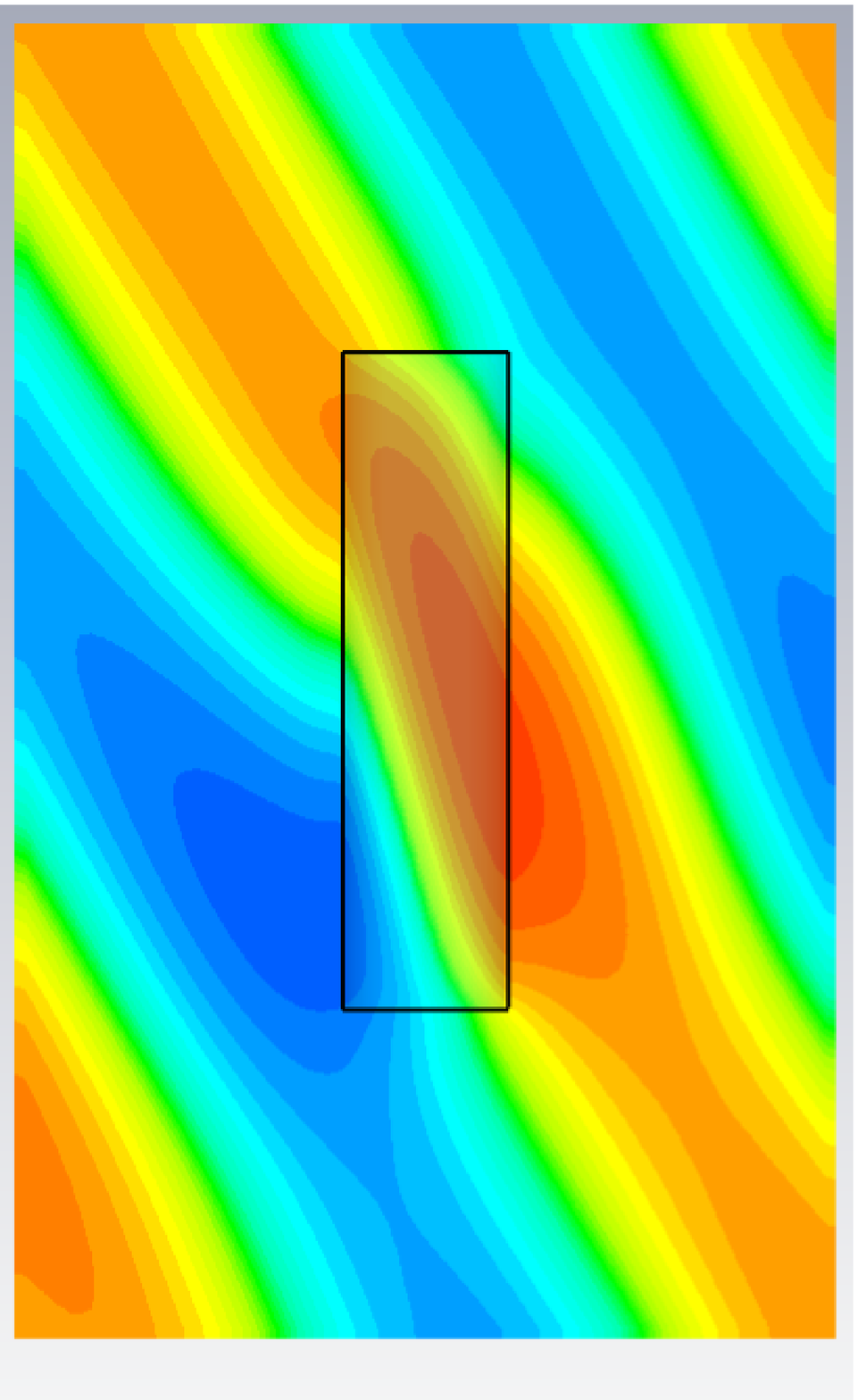}
\includegraphics[width=9.4cm]{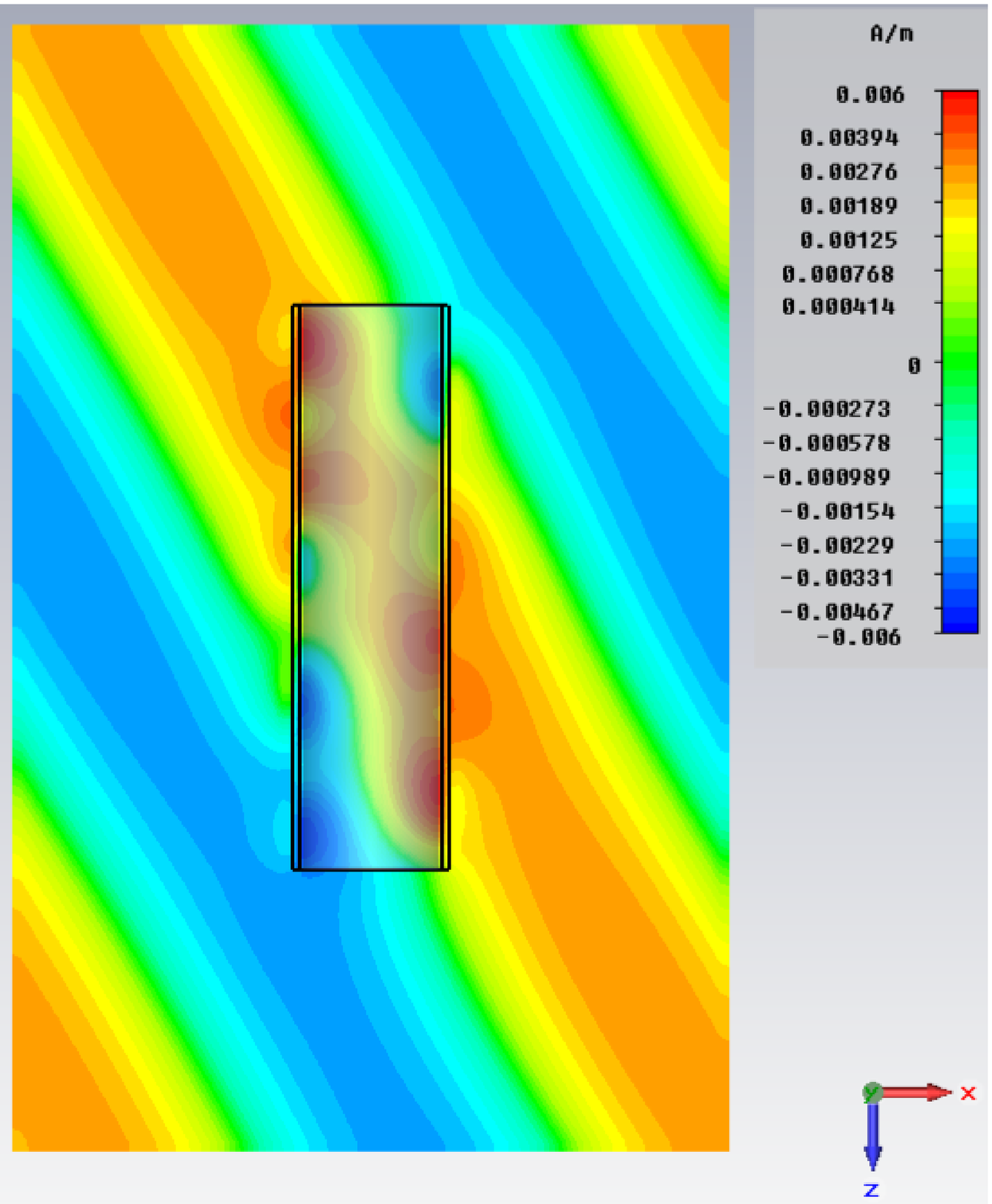}
\caption{Contour plots (snapshot in time) of the magnetic field on the
  $E$ plane of polarization for a TM$_z$ wave at $\alpha=\pi/3$
  (origin top right corner) for the object of
  Fig.~\protect\ref{fig:UvC}.  (a) Uncloaked; (b) the $a_c=1.1a$
  cloak.  Severe uncloaked wavefront distortions are almost totally
  restored by the thin cloak.  Cloak interface nodes (right) are
  associated with plasmonic surface waves.}
\label{fig:H-field}
\vspace*{-5mm}
\end{figure}

\begin{figure}[!pb]
(a) uncloaked cylinder
\qquad\qquad\qquad\quad
(b) cloaked cylinder
\qquad\qquad\qquad\quad
(c) cloaked cylinder (zoomed in) \\[2mm]
\centering
\includegraphics[width=4.95cm]{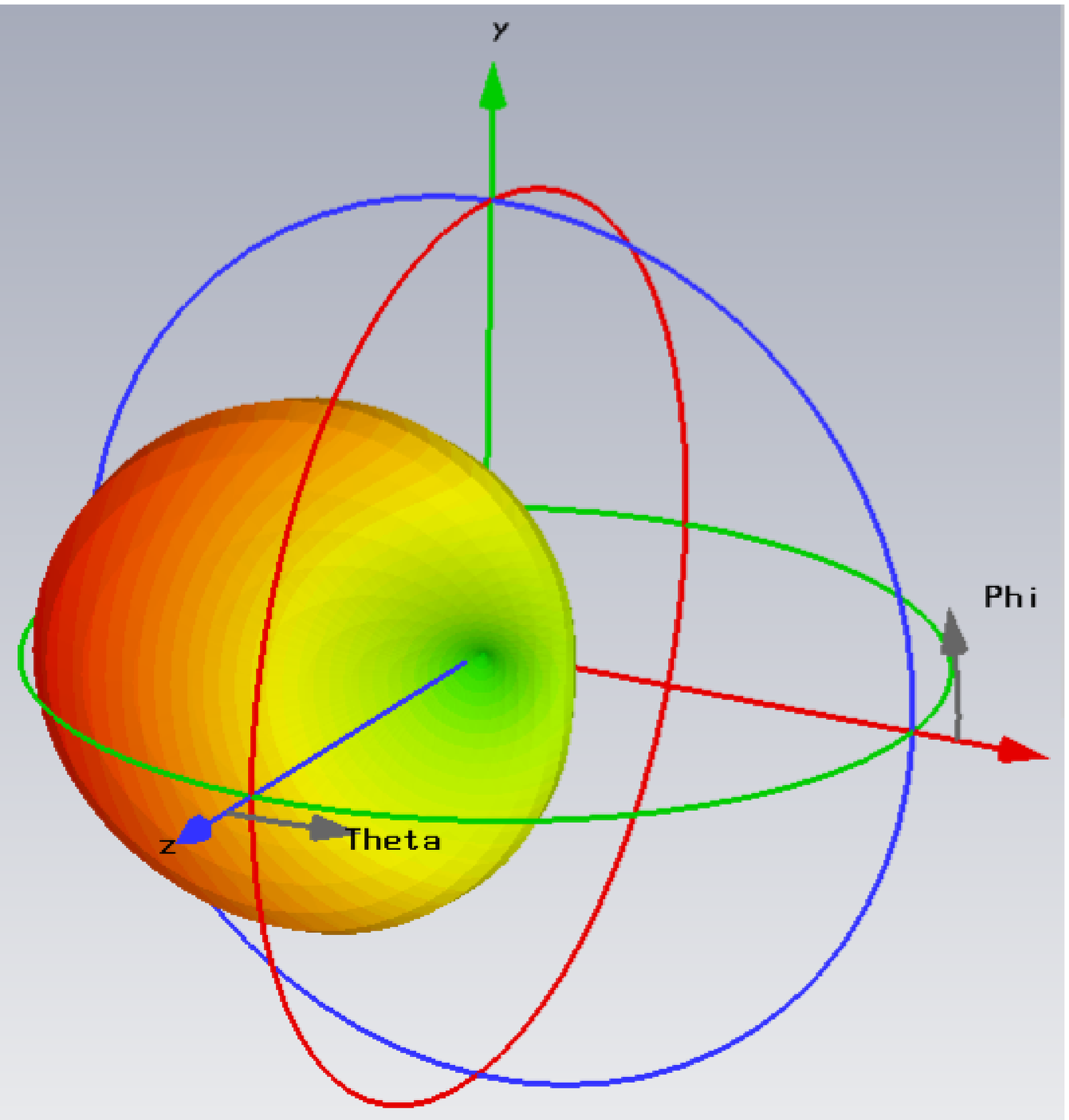}
\includegraphics[width=5.95cm]{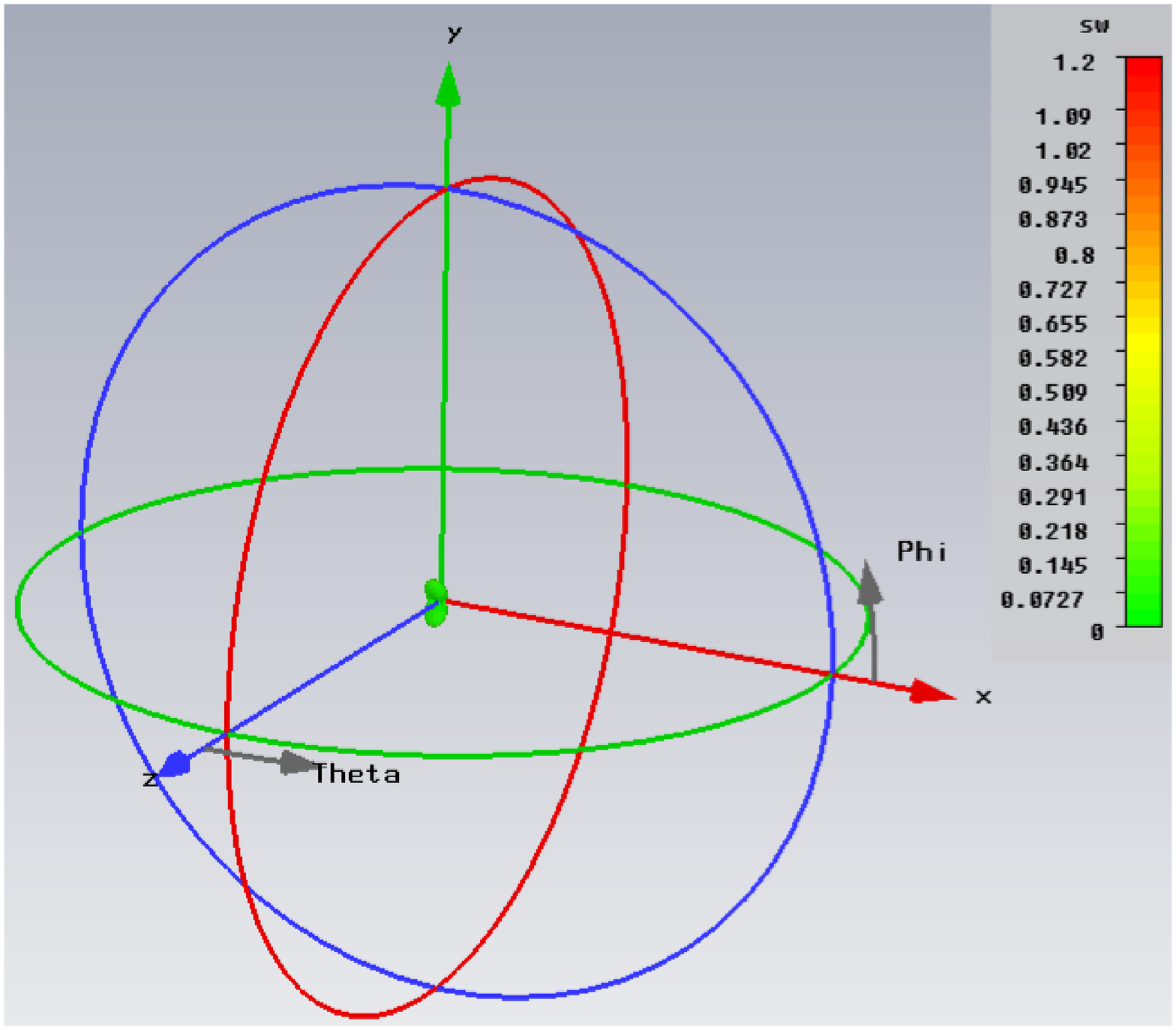}
\includegraphics[width=5.9cm]{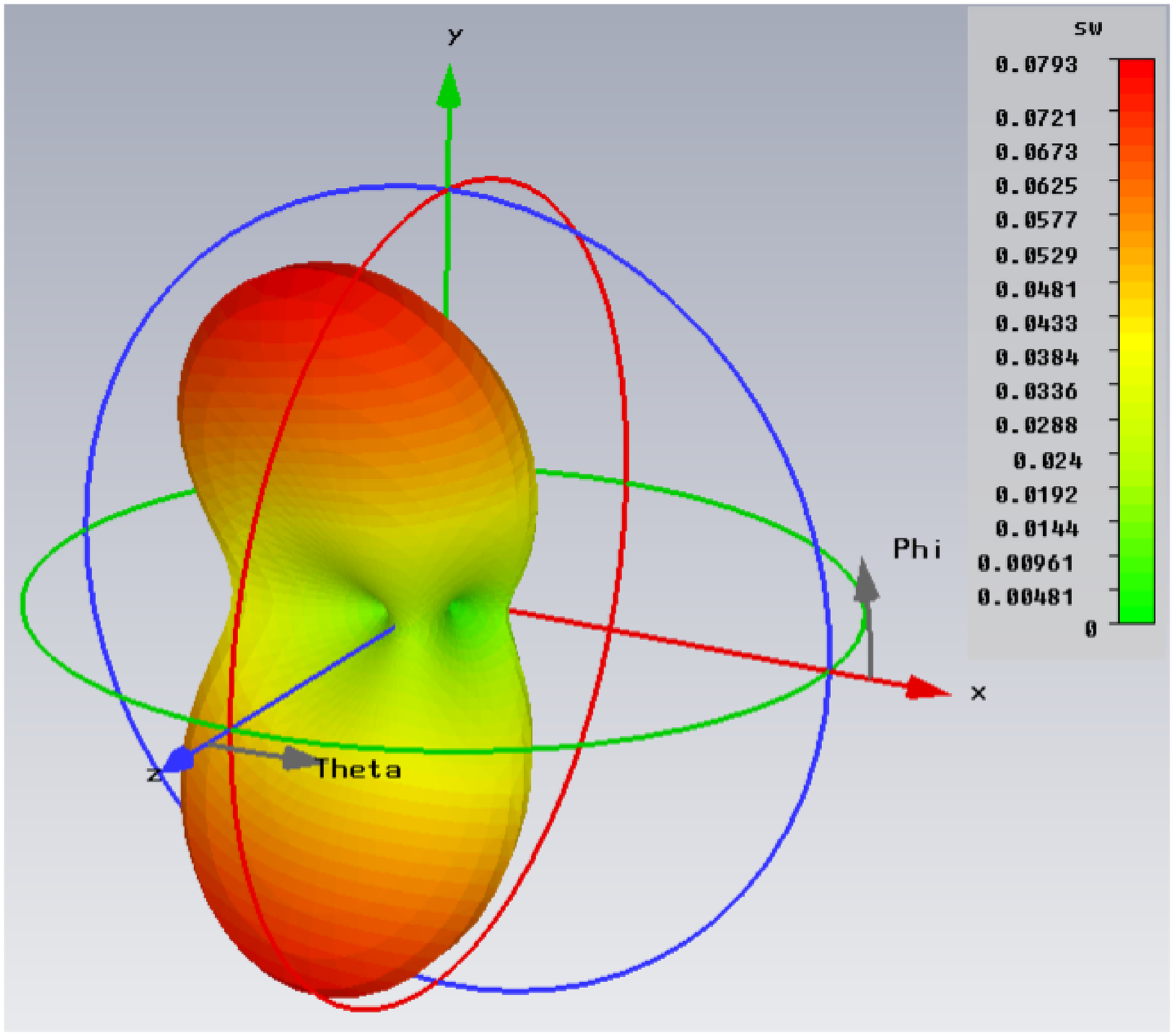}
\caption{Far-field scattering patterns in the (a) uncloaked and (b)
  cloaked ($a_c=1.1a$) cylinders of Fig.~\protect\ref{fig:UvC},
  plotted on the same scale; (c) is an enlargement of (b), to show the
  residual detail of the scattering pattern, dominated by higher-order
  scattering modes.  Panels (a) and (b) demonstrate the dramatic
  scattering reduction from a properly tuned plasmonic cloak.}
\label{fig:farfield}
\end{figure}

Figs.~\ref{fig:H-field}-\ref{fig:farfield} illustrate the full-wave
near- and far-field numerical results for the cylinder of
Sec.~\ref{sub:primary} (Fig.~\ref{fig:UvC}), illuminated by a TM wave
impinging at an angle $\alpha=\pi/3$.  The various figures compare in
panel (b) the cloaked configuration at the design frequency $f_0$ with
the uncloaked case (a).  Consistent with Fig.~\ref{fig:th-oblique}a,
the overall calculated RCS reduction at the central frequency is
-14.2~dB, and the following figures depict the effective functionality
of the cloak in its near- and far-field regions.  We have chosen
$\alpha=\pi/3$ to indicate the performance of the cloak to oblique
excitation; similar considerations apply for any other angle of
incidence, which we have verified via detailed study.

Fig.~\ref{fig:H-field} shows the (normal) magnetic field distribution
on the E plane (snapshot in time), comparing the uncloaked scenario
(a) with the cloaked one (b).  Several interesting features are
noteworthy.  In the uncloaked case, the wave penetrates the dielectric
rod and experiences a wavelength shortening that effectively distorts
the planar wave fronts on the back of the cylinder, producing
significant shadow and scattered fields all around the object.  The
thin plasmonic metamaterial shell is able to re-establish the proper
planar fronts just outside the cloak (b), ensuring reduced scattering
and suppressed visibility for an outside observer positioned anywhere
in the near- or far-field of the object.  It should be noted that this
effect is obtained for oblique incidence, and the cylinder ends are
uncloaked.  Additional improvement may be achieved by proper cloaking
of these terminations.  We may also easily observe how the plasmonic
layer supports surface-plasmon waves traveling along the shell, as
expected due to its negative permittivity.  It is interesting to
observe how these waves effectively cancel the residual scattering and
restore the phase fronts to almost exactly match those of the original
plane wave, had it traveled through free space instead.

Fig.~\ref{fig:farfield} illustrates the far-field radiation patterns
for this same cylinder, comparing on the same scale scattering from
the uncloaked (a) and cloaked (b) objects, showing drastic suppression
of the bistatic RCS at all angles.  Uncloaked scattering exhibits
itself mainly as a shadow on the cylinder backside; various
higher-order scattering harmonics contribute to this residual
scattering pattern.  Most of the scattering is suppressed by the
cloak.  Panel (c) shows the cloaked residual scattering on a much
smaller scale: as expected, scattering is not identically zero, and
small lobes, associated with higher-order (and more directive)
cylindrical harmonics not completely suppressed, are still present.
Their relevance, however, is very limited compared to the original
scattering levels.

%


\section{Conclusions}

We have presented an extensive investigation of the application of the
plasmonic cloaking technique to circular cylinders illuminated by
plane waves of arbitrary polarization and angle of incidence.  We have
derived analytical formulas for the general oblique-angle scenario,
and showed that in the electrically-thin limit there is no angular
dependence on the cloaking response, i.e.\ the design formulas are
independent on the angle of incidence.

To study the characteristics of cylindrical cloaks, we have designed a
set of TM$_z$-optimized cloaks for a wide range of parameters and
cylinders of interest, using the normal-incidence analytical formulas
with realistic losses and metamaterial frequency dispersion
implemented in a Drude model.  For the cloak design, we have focused
on TM polarization because it dominates the scattering of moderately
thick dielectric and conducting cylinders, which is of interest for
several applications within the radar community.  The optimized
cloaks' drastic scattering reduction was corroborated with full-wave
numerical simulation, taking into account variations of the angle of
incidence, core permittivity, cylinder diameter, and most importantly
truncation effects due to finite length.

We have found that, as predicted by the analytical formulas presented
here, for elongated objects with diameters up to one-half the
wavelength (for which our cloaking technique is most effective), but
with length comparable with the wavelength of operation, a simple
one-layer permittivity cloak is very effective, providing significant
scattering reduction highly robust to variations in the angle of
incidence.  Performance is slightly weakened at near-grazing angles,
for which TM--TE polarization coupling partially affects the overall
performance of a single-layer cloak optimized for TM polarization
alone.  We note, however, that scattering approaching grazing angles
is the weakest in absolute value, thus less important for achieving
overall scattering reduction.

The analytical theory developed here for infinite cylinders has been
strongly corroborated by our numerical simulations for finite lengths,
implying that the truncation effects do not significantly perturb the
cloaking effect.  We are currently working on the extension of these
concepts to multi-layered metamaterial cloaks for suppression of
multiple scattering coefficients, and contemporary suppression of TE
and TM scattered waves, which may increase the size of the cloaked
objects and the angular range of operation. We are also currently
pursuing an experimental realization of the plasmonic cloaking concept
at radio frequencies for practical finite cylinders and cloaks.


\appendix*


\section{Analytical Solution of the Scattering from a Two-Layer Circular Cylinder for Oblique Incidence}
\label{app}

Consider the geometry of Fig.~\ref{fig:cyl}, consisting of a circular
cylinder of radius $a$, permittivity $\epsilon$ and permeability
$\mu$, covered by a thin conformal cylindrical cloak shell of
thickness $(a_c-a)$, permittivity $\epsilon_c$ and permeability
$\mu_c$.  When excited by an impinging infinite plane wave of
arbitrary incidence angle and polarization, the general scattering
problem may be solved in the limit of long cylinders ($L\to\infty$) by
expanding the impinging and scattered fields in terms of cylindrical
harmonics~\cite{Balanis1989,Bohren1983,Yousif1994}.  The problem may
be split into the orthogonal polarizations with $E$-field transverse
to the cylinder axis (TE$_z$) and $H$-field transverse (TM$_z$).

For a TM$_z$ plane wave impinging at an angle $\alpha$ from the
cylinder axis $\hat{z}$, the incoming magnetic field may be written
as:
\bq\label{eq:H-field}
H_i \, = \,
\frac{E_0}{\eta_0} \, e^{-i\beta z} \, e^{-ik_0^Tx} \, \hat{x}
\; ,
\eq
where $E_0$ is the electric field amplitude, $\eta_0$ and $k_0$ are
the free-space characteristic impedance and wave number,
$\beta=k_0\cos\alpha$ is the wave number component along the cylinder
axis, $k_0^T=\sqrt{k_0^2-\beta^2}=k_0\sin\alpha$ is the transverse
component of the wave number, and the corresponding impinging electric
field may be calculated using the curl Maxwell's equations.  Expanding
in cylindrical waves, we may write electric and magnetic fields as:
%
\ba\notag
E_{\rm TM} \, = \, &
 \frac{1}{(k_0^T)^2}
 \frac{\partial^2 u_i^{\rm TM}}{\partial\rho\partial z}
 \hat\rho & + \;
 \frac{1}{\rho (k_0^T)^2}
 \frac{\partial^2 u_i^{\rm TM}}{\partial\phi\partial z}
 \hat\rho
 \; + \;
 u_i^{\rm TM}\hat{z}
\; , \\ \label{eq:EH}
H_{\rm TM} \, = \, &
 \frac{ik_0}{\rho \eta_0 (k_0^T)^2}
 \frac{\partial u_i^{\rm TM}}{\partial\phi}
 \hat\rho & - \;
 \frac{ik_0}{\rho (k_0^T)^2}
 \frac{\partial u_i^{\rm TM}}{\partial\rho}
 \hat\phi
\; ,
\ea
where
\bq\label{eq:uiTM}
u_i^{\rm TM} \, = \, E_0 \, i^{-n} (\sin\alpha) \, J_n(k_0^T\rho)
                     \, e^{in\phi} \, e^{-i\beta z}
\eq
and $J_n$ is the cylindrical Bessel function of order $n$.  Analogous
expressions for TE$_z$ waves may be derived using duality.

Using the orthogonality of cylindrical waves, the boundary conditions
at the radial interfaces may be met by assuming the existence of
transmitted cylindrical waves in the two dielectric regions and a
scattered wave in free-space, which may be written consistently with
Eq.~(\ref{eq:EH}), using the scalar potentials:

\ba\notag
u_1^{\rm TM} \, = \, 
i^{-n} E_0 \, e^{in\phi} \, e^{-i\beta z} \sin\alpha &
c_{1,n}^{\rm TM} J_n(k_1^T\rho)
& \quad \rho < a \, ,
\\ \label{eq:pot}
u_2^{\rm TM} \, = \, 
i^{-n} E_0 \, e^{in\phi} \, e^{-i\beta z} \sin\alpha &
[  c_{2i,n}^{\rm TM} J_n(k_2^T\rho)
 + c_{2o,n}^{\rm TM} Y_n(k_2^T\rho) ]
& \quad a < \rho < a_c \, ,
\\ \notag
u_s^{\rm TM} \, = \, 
i^{-n} E_0 \, e^{in\phi} \, e^{-i\beta z} \sin\alpha &
c_{s,n}^{\rm TM} H_n^{(1)}(k_0^T\rho)
& \quad \rho > a_c \, ,
\ea
for the fields induced in the core region, shell, and for the the
scattered field, respectively.  The core and cloak regions are labeled
``1'' and ``2'', respectively, and ``s'' represents the scattered wave
outside the cloak; $k_i$ are the relevant transverse wave numbers for
each region and $k^T_i=\sqrt{k_i^2-\beta^2}$.  $J_n$ and $Y_n$ are the
cylindrical Bessel functions of the first and second kind, for
incoming and outgoing waves, while $H_n^{(1)}$ is the cylindrical
Hankel function.  The complex scattering coefficients are the
unknowns.  Analogous equations may be written for TE$_z$ waves.

For normal incidence ($\alpha=\pi/2$), for PEC objects, or for the
azimuthally symmetric mode $n=0$, the problem is easily solved by
matching the two non-zero tangential field components at each radial
interface:
\ba\notag
E_{s,z(\phi)} \, = \, & E_{2,z(\phi)} & \quad \rho = a_c \, , \\ \notag
E_{2,z(\phi)} \, = \, & E_{1,z(\phi)} & \quad \rho = a   \, , \\ \notag
H_{s,\phi(z)} \, = \, & H_{2,\phi(z)} & \quad \rho = a_c \, , \\ \label{eq:BCs}
H_{2,\phi(z)} \, = \, & H_{1,\phi(z)} & \quad \rho = a   \, ,
\ea
for TM$_z$(TE$_z$) polarization.  This yields the familiar rank-four
determinant expressions reported in Eqs.~(\ref{eq:Un},\ref{eq:Vn}).
Consistent expressions for the TE$_z$ coefficients may be obtained by
applying duality.  In the general case of oblique incidence on a
dielectric cylinder, however, the higher-order $n>0$ modes are
characterized by all four independent tangential components of the
fields at each interface, which may not be matched independently for
each TE$_z$ or TM$_z$ harmonic.  The boundary conditions may instead
be met, as derived in~\cite{Yousif1994}, by linearly combining TE$_z$
and TM$_z$ harmonics of same order $n>0$.  The corresponding solution
of an eight-by-eight system of equations, derived
in~\cite{Yousif1994}, provides the exact expression of the scattering
coefficients in the general case, to be used in Eq.(~\ref{eq:sum}) to
derive the total scattering width of the cylinder.  This form of
polarization coupling is inherently associated with the asymmetry
introduced by the oblique excitation for higher-order modes, and it
may be avoided only in the case of conducting objects or for $n=0$.
In the quasi-static limit, however, the dependence of the arguments of
the Bessel functions in Eqs.~\ref{eq:pot} on the transverse wave
number $k_c$ is negligible as well; therefore this form of
cross-polarization coupling is negligible.  In this limit,
Eqs.~(\ref{eq:Un}-\ref{eq:Vn}) may be used for any angle of incidence,
at the basis of the derivation of
Eqs.~(\ref{eq:FOT-gen}-\ref{eq:FOT-PEC}), which do not depend on
$\alpha$.


\begin{acknowledgments}

A.~A.\ was partially supported by the National Science Foundation
(NSF) CAREER award ECCS-0953311.  A.~K.\ and D.~R.\ were supported by
an internal research award at ARL:UT.

\end{acknowledgments}



\end{document}